\documentclass{article}

\usepackage[a4paper, margin=2.5cm]{geometry} 
\usepackage{parskip}  
\usepackage{authblk}  
\usepackage[dvipsnames]{xcolor}   
\usepackage[style=chem-acs]{biblatex} 
\usepackage{bm}       
\usepackage{amssymb}  
\usepackage{amsmath}  
\usepackage{hyperref} 
\usepackage{graphicx} 
\usepackage{braket}   
\usepackage{booktabs} 
\usepackage{cleveref} 
\usepackage{setspace} 
\usepackage[version=4]{mhchem}   

\DeclareMathOperator*{\argmin}{arg min}
\newcommand{\updownarrows}{\mathbin\uparrow\hspace{-.4em}\downarrow}
\newcommand{\br}{\mathbf{r}}
\newcommand{\bR}{\mathbf{R}}
\newcommand{\bs}{\bm{\sigma}}
\newcommand{\bx}{\mathbf{x}}
\newcommand{\bh}{\mathbf{h}}
\newcommand{\bv}{\mathbf{v}}
\newcommand{\bw}{\mathbf{w}}
\newcommand{\bk}{\mathbf{k}}
\newcommand{\bc}{\mathbf{c}}
\newcommand{\bp}{\mathbf{p}}
\newcommand{\bS}{\mathbf{S}}
\newcommand{\bH}{\mathbf{H}}
\newcommand{\bK}{\mathbf{K}}
\newcommand{\dr}{\mathrm{d}\mathbf{r}}

\title{Machine Learning Wavefunction}
\author{Stefano Battaglia
\thanks{stefano.battaglia@kemi.uu.se}}
\affil{Department of Chemistry - BMC, Uppsala University, SE-75123 Uppsala, Sweden}

\begin{document}

\doublespacing

\maketitle

\begin{abstract}
This chapter introduces the main ideas and the most important methods for
representing the electronic wavefunction through machine learning models.
The wavefunction of a $N$-electron system is an incredibly complicated
mathematical object, and models thereof require enough flexibility
to properly describe the complex interactions between the particles, but
at the same time a sufficiently compact representation to be useful in
practice.
Machine learning techniques offer an ideal mathematical framework to
satisfy these requirements, and provide algorithms for their optimization
in both supervised and unsupervised fashions.
In this chapter, various examples of machine learning wavefunctions are
presented and their strengths and weaknesses with respect to traditional
quantum chemical approaches are discussed; first in theory, and then in
practice with two case studies.
\end{abstract}

\section{Introduction}

We shall start this chapter by introducing the mathematical infrastructure in which
wavefunction methods are defined, and then provide a brief overview of the machine
learning (ML) approaches that have been developed within this framework.
This will set the stage for the remainder of this chapter, where we will discuss
in more detail a number of methods that successfully leverage ML techniques to
represent the electronic wavefunction.

\begin{figure}
  \centering
  \includegraphics[width=10cm]{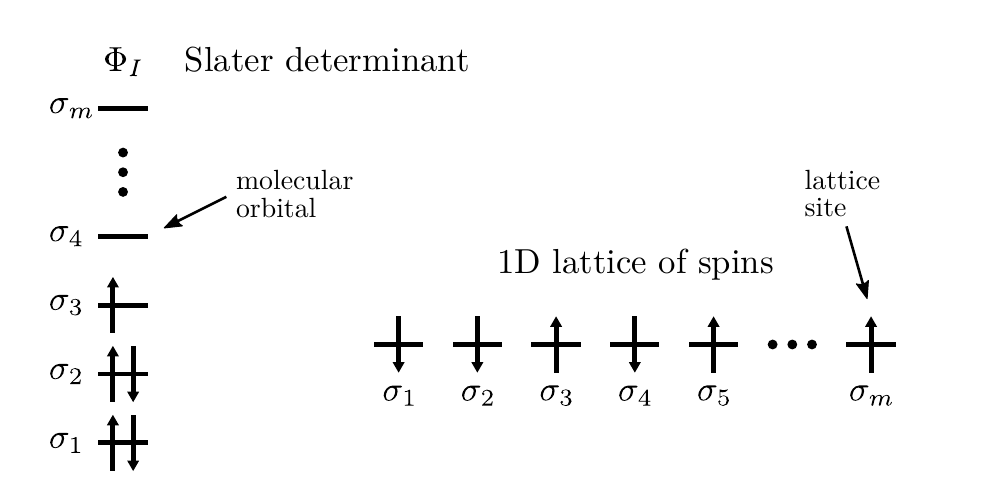}
  \caption{Graphical representation of a Slater determinant (left) and of
  a 1D lattice (right). The occupation numbers $\sigma_i$ refer to molecular
  orbitals for the SD, and to individual spins on lattice sites for
  the 1D lattice.}
  \label{fig:configurations}
\end{figure}

At the heart of wavefunction theory and the electronic structure problem lies the
time-independent Schrödinger equation (TISE)
\begin{equation}
  \label{eq:TISE}
  \hat{H} \Psi = E \Psi
\end{equation}
where $\Psi$ is the wavefunction describing the quantum state, $E$ is its
associated energy, and $\hat{H}$ is the \textit{ab initio} electronic Hamiltonian.
The latter is given (in atomic units) by
\begin{equation}
  \label{eq:H_electronic}
  \hat{H} = - \sum_{i=1}^N \frac{1}{2}\nabla_i
  - \sum_{i=1}^N \sum_{I=1}^{N_{atoms}} \frac{Z_I}{|\mathbf{r}_i - \mathbf{R}_I|}
  + \sum_{i=1}^N \sum_{j>i}^N \frac{1}{|\mathbf{r}_i - \mathbf{r}_j|}
  + \sum_{I=1}^{N_{atoms}} \sum_{J>I}^{N_{atoms}} \frac{Z_I Z_J}{|\mathbf{R}_I - \mathbf{R}_J|}
\end{equation}
with the first term describing the kinetic energy of the electrons, the second term
the attraction between electrons and nuclei, and the third and fourth terms the
electron-electron and nuclear-nuclear repulsion, respectively.
Within the Born\textendash Oppenheimer approximation, the wavefunction $\Psi$ only
depends on the position $\br = (\br_1,\ldots,\br_N)$ of the $N$ electrons
\begin{equation}
  \label{eq:WF}
  \Psi \equiv \Psi(\br_1,\ldots,\br_N) = \Psi(\br)
\end{equation}
and encodes all the information of the $N$-particle system.
This mathematical object is incredibly complex and constitutes the quantity subject to
approximations in wavefunction theory.
Typically, the first step in the practical resolution of \Cref{eq:TISE} is the introduction
of a finite many-particle basis, $\{\Phi_I\}_{I=0}^{M}$, and the expansion of the
wavefunction $\Psi$ in this basis.
We thus have the following \emph{ansatz}
\begin{equation}
  \label{eq:FCI}
  \Psi(\mathbf{r}_1,\ldots,\mathbf{r}_N) = \sum_{I=0}^{M}
  C_I \Phi_I(\mathbf{r}_1,\ldots,\mathbf{r}_N)
\end{equation}
where the coefficients $C_I$ are \textit{a priori} unknown parameters (forming
a vector $\mathbf{C}$), that can be determined by minimizing the energy
expectation value according to the variational principle
\begin{equation}
  \label{eq:Evar}
  \mathbf{C}^* = \argmin_{\mathbf{C}}
  \frac{\braket{\Psi|\hat{H}|\Psi}}{\braket{\Psi|\Psi}}
\end{equation}
Not \emph{any} type of many-particle function $\Phi_I$ can be used in \Cref{eq:FCI},
because a \emph{fermionic} wavefunction has to be antisymmetric with respect
to the permutations of two identical particles, \textit{i.e.}
\begin{equation}
  \label{eq:antisymmetry}
  \Psi(\mathbf{r}_1,\ldots,\mathbf{r}_i,\mathbf{r}_j,\ldots,\mathbf{r}_N) =
  - \Psi(\mathbf{r}_1,\ldots,\mathbf{r}_j,\mathbf{r}_i,\ldots,\mathbf{r}_N)
\end{equation}
The most common and practical choice to explicitly enforce this property is to choose
Slater determinants (SDs) as the many-particle functions $\Phi_I$, which are
constructed from a finite one-electron basis $\{\phi_i\}_{i=1}^{m}$.
Importantly, the introduction of this orbital basis allows us to express \Cref{eq:FCI}
as a linear combination of occupation number vectors in \emph{Fock space}
\begin{equation}
  \label{eq:FCI_ONV}
  \ket{\Psi} = \sum_{\sigma_1,\sigma_2,\ldots,\sigma_m}
  \psi_{\sigma_1,\sigma_2,\ldots,\sigma_m}
  \ket{\sigma_1,\sigma_2,\ldots,\sigma_m}
  = \sum_{\bm{\sigma}}\psi_{\bm{\sigma}}\ket{\bm{\sigma}}
\end{equation}
where $\sigma_i = 0$ if the \emph{spin-orbital} $\phi_i$ is empty and $\sigma_i = 1$
if it is occupied (note that it is also common to use \emph{molecular orbitals},
in which case we would have 4 possible states:
$\sigma_i \in \{\cdot,\uparrow,\downarrow,\updownarrows\}$).
In this representation, a Slater determinant $\Phi_I$ is encoded by the occupation
pattern of the orbitals, and the \emph{amplitudes} $\psi_{\bm{\sigma}}$ parametrizing
the wavefunction, are labeled by the corresponding \emph{many-body configuration}
$\bm{\sigma} = (\sigma_1,\sigma_2,\ldots,\sigma_m)$.
Note that the amplitudes $\psi_{\bs}$ in \Cref{eq:FCI_ONV} are nothing but a relabeling
of the coefficients $C_I$ of \Cref{eq:FCI}.
One advantage of \Cref{eq:FCI_ONV} is that it also naturally represents quantum states
of model Hamiltonians that are defined on discrete lattices, see \Cref{fig:configurations}
for the comparison of a Slater determinant and a lattice.
These are the systems targeted by the first machine learning approaches
that we will see in the next section, thus it is important to draw the connection
between them and the language of quantum chemists.
When the many-body configurations in \Cref{eq:FCI_ONV} represent Slater determinants,
the ansatz is called full configuration interaction (FCI).
Within a given one-particle basis, the FCI method yields the exact solution of
\Cref{eq:TISE}, however it is only applicable to few-particle systems due to its
unfavorable computational scaling.
In fact, the FCI ansatz highlights one major challenge intrinsic to the quantum many-body
problem, that is, the size of the space of many-body configurations in which the solution
$\Psi$ of \Cref{eq:TISE} lives in, grows exponentially with the number of particles $N$
(or, equivalently, the number of one-particle basis functions $m$).
This is manifestly visible in \Cref{eq:FCI_ONV}, where the sum runs over all possible
$2^m$ many-body configurations.
A major goal of electronic structure theory is to find approximations to \Cref{eq:FCI_ONV}
that only contain a polynomial number of parameters, yet capturing the most important
correlation features by retaining only the dominant configurations.
To this end, many approximate wave function ansätze have been proposed throughout
the years.
Historically, the predominant recipe to approximate wavefunctions has been to truncate
the sum in \Cref{eq:FCI_ONV} according to the number of excitations from a reference
SD (or several reference SDs), resulting in a series of systematically improvable ansätze.
These are for example the configuration interaction, coupled cluster and perturbation
theory approaches discussed briefly in chapter 1.
More recently, alternative parametrizations stemming from the condensed matter physics
community have emerged, such as matrix product states, where the idea is to fix a maximum
number of many-particle functions, but allow them to change iteratively in an algorithm
known as the density matrix renormalization group \cite{White1992}.
If only a single many-particle basis function is retained in the ansatz of
\Cref{eq:FCI}, we recover the Hartree-Fock approximation, a cornerstone of electronic
structure theory since its inception. From this discussion, it is clear that
the FCI ansatz offers the greatest flexibility, however it is computationally untractable
due to its exponential scaling.
On the other end of the spectrum, the HF approximation is the most compact, but its
accuracy is not sufficient for most applications.
We need to find the right compromise between these two extremes; this is the entry
point for machine learning approaches.

The astonishing success that machine learning is having in representing
high-dimensional data with complex dependencies offers new avenues for the electronic
structure problem. In fact, machine learning techniques to solve the Schrödinger equation
were already explored more than a decade ago
\cite{Lagaris1997,Sugawara2001,Manzhos2009,Caetano2011}, despite passing relatively unnoticed.
It is only thanks to the current success that ML is having in other fields of computational
chemistry and physics, that the interest to apply it to wavefunction theory has resurged
as well.
There are several ways to harness the power and flexibility of machine learning
in the context of wavefunction theory.
One possibility is to use them to improve or accelerate existing methods, such
as in the machine learning configuration interaction approach by \textcite{Coe2018}
or the accelerated coupled cluster by \textcite{Townsend2019}.
These ideas and their implementation details were the subject of chapter 22
and will not be discussed further here.
Instead, the main topic for the remainder of this chapter are ML models that
are \emph{directly} used to represent the wavefunction. These try to completely
bypass the standard framework of expressing an analytically integrable ansatz
(typically based on Gaussian orbitals), and obtain the optimal parameters through the
solution of an eigenvalue equation or many-body conditions.
In this respect, the very first work in recent times that showed the true power and
flexibility of ML approaches to solve the Schrödinger equation is that by
\textcite{Carleo2017}. Here, the authors proposed a particular type of
neural network (NN) --- the restricted Boltzmann machine (RBM) --- that was able
to encode the ground state of two paradigmatic spin Hamiltonians with state-of-the-art
accuracy. The exceptional result of this \emph{neural-network quantum state} (NQS)
ansatz led to a large number of follow-up works within the condensed matter physics
community.
For instance, the relation of RBMs to the more known tensor network states was
quickly made \cite{Chen2018,Clark2018,Glasser2018,Li2021b,Huang2021} and their
characterization in terms of quantum entanglement \cite{Deng2017a} and
representability theorems \cite{Gao2017} was established. It was shown how to
include abelian and non-abelian symmetries in the ansatz \cite{Choo2018,Vieijra2020},
meanwhile many extensions of, and alternatives to, the simple RBM architecture to more
general NNs were developed, resulting in new NQS ansätze
\cite{Nomura2017,Xia2018,Carleo2018,Cai2018,Liang2018,Luo2019,Hibat-Allah2020,Inui2021}.
The application of restricted Boltzmann machines to solve the TISE with the
full \textit{ab initio} electronic Hamiltonian was also shown possible.
In a few cases, the electronic structure problem was mapped from fermionic degrees
of freedom to spin ones, and essentially the same technique as used in the
original work by \textcite{Carleo2017} was then used to obtain the ground state
energy of several small molecular systems and the dissociation curves of a few
diatomic molecules \cite{Xia2018,Choo2020}. The accuracy reached in this case was on-par
or beyond that of CCSD(T), albeit only in conjunction with a minimal basis set.
Neural-network quantum states have also seen application as active space solvers
in the context of the CASSCF method, with promising results \cite{Yang2020}.
A significant step forward was made with the development of two similar ansätze
based on deep neural networks (DNNs). The \emph{FermiNet} \cite{Pfau2020} and
\emph{PauliNet} \cite{Hermann2020} architectures completely bypass the typical
dependence on a one-electron basis set and directly represent the wavefunction in
real space. While an earlier attempt based on DNNs did not consistently reach
an acceptable accuracy \cite{Han2019}, both FermiNet and PauliNet yielded results
comparable or superior to the best methods currently available across the board.
All the ML-based wavefunctions mentioned so far are true \textit{ab initio}
ansätze in the sense that do not require prior data to be trained, rather
their parameters are optimized in an unsupervised fashion.
On the other hand, supervised techniques are also possible.
Examples of these have been already discussed in chapter 18 in the context
of DFT, where for instance, the electronic density was learned from reference
data and predicted by Gaussian process regression \cite{Grisafi2019}
or neural networks \cite{Chandrasekaran2019}.
In the same spirit, but in the framework of wavefunction theory,
the \emph{SchNOrb} deep convolutional neural network predicts the electronic wavefunction
by learning the Hamiltonian and overlap matrices expressed in the molecular orbital
basis from a set of reference calculations \cite{Schutt2019,Gastegger2020}.\\
At last, while neural networks have been clearly the favorite choice
so far, recent works exploring the efficacy of Gaussian processes to represent
the wavefunction have shown that non-parametric approaches are equally valid
alternatives \cite{Glielmo2020,Rath2020}.
In fact, the so-called \emph{Gaussian process state (GPS)} is able to reach
and surpass the accuracy of RBMs in the solution of the Fermi-Hubbard model,
with a very compact representation of the ground state wavefunction.

In the next section we shall discuss in more detail a number of different ML
approaches to the solution of the Schrödinger equation and place them
in the larger context of electronic structure theory.
We shall analyze strengths and weaknesses of these methods and see how they
compare to their traditional counterparts.
Finally, you will have the chance to get first-hand experience with these new
powerful tools through two case studies.

\section{Methods}

The discussion of this section is divided based on the (mathematical)
space in which the methods are defined.
First, we look at approaches expressed in the Fock space of many-body
configurations, as these are conceptually closer to the usual framework
used in traditional quantum chemical methods.
Second, we consider ansätze in first quantization, that is, defined directly
in the real space of electronic coordinates.
The third subsection is devoted to a supervised method that is neither defined
in Fock space, nor in real space. Instead, it infers the wavefunction of a
system directly from the molecular geometry.
However, before dwelling into the discussion of these methodologies, we shall
go through a short introduction to variational Monte Carlo (VMC), as this
technique is a common denominator for the complicated ansätze considered here.

\subsection{Variational Monte Carlo in a nutshell}

The underlying idea of variational Monte Carlo is to evaluate the
high-dimensional integrals appearing in the quantum many-body problem by the
Monte Carlo method \cite{Toulouse2016}.
Consider the energy expectation value $E$ associated to the quantum state $\Psi$,
\begin{equation}
  \label{eq:E_expval}
  E = \frac{\braket{\Psi|\hat{H}|\Psi}}{\braket{\Psi|\Psi}}
  = \frac{\int \Psi^*(\br)\hat{H}\Psi(\br)\dr}{\int \Psi^*(\br)\Psi(\br)\dr}
\end{equation}
The right-hand side contains integrals over the whole $3N$-dimensional space
which are hard to evaluate numerically. However, the integrands can be
manipulated into more a convenient expression which is amenable for
the Monte Carlo technique. That is,
\begin{equation}
  \label{eq:E_expval_VMC}
  \frac{\int \Psi^*(\br)\hat{H}\Psi(\br)\dr}{\int \Psi^*(\br)\Psi(\br)\dr}
  = \frac{\int |\Psi(\br)|^2 \tfrac{\hat{H}\Psi(\br)}{\Psi(\br)} \dr}
    {\int |\Psi(\br)|^2 \dr}
  = \int \frac{|\Psi(\br)|^2}{\int |\Psi(\br)|^2 \dr}
  \frac{\hat{H}\Psi(\br)}{\Psi(\br)} \dr
  = \int \rho(\br) E_{loc}(\br) \dr
\end{equation}
where $\rho(\br)$ is interpreted as a probability distribution and
$E_{loc}(\br) = \tfrac{\hat{H}\Psi(\br)}{\Psi(\br)}$ is the so-called \emph{local energy}.
By drawing a finite number $N_{s}$ of sample points $\{\br^{(k)}\}_{k=1}^{N_s}$
according to $\rho(\br)$,
the energy expectation value can be approximated as an average over local energies as
\begin{equation}
  \label{eq:VMC_energy}
  E \approx \frac{1}{N_{s}} \sum_{k=1}^{N_{s}} E_{loc}(\br^{(k)})
\end{equation}
A typical choice to sample the points $\{\br^{(k)}\}_{k=1}^{N_s}$ is to walk in
$3N$-dimensional space and generate a Markov chain of electronic coordinates,
$\br^{(1)}\to\br^{(2)}\to\ldots\to\br^{(N_s)}$, through the Metropolis-Hastings
algorithm \cite{Metropolis1953,Hastings1970}. This is an efficient way to generate
the new positions, because the usually complicated integral in the denominator of
$\rho(\br)$ does not need to be evaluated.
The traditional wavefunction ansatz used in VMC
(also called trial wavefunction within the quantum Monte Carlo community)
is of the \emph{Slater-Jastrow} type \cite{Jastrow1955}, and has the following
general form
\begin{equation}
  \label{eq:SJ}
  \Psi_{\bm\Theta}(\br) = \mathcal{J}(\br,\bm\Theta)\Phi_0(\br)
\end{equation}
Here, $\Phi_0(\br)$ is a mean-field solution (such as the Hartree-Fock determinant)
or a small linear combination of SDs, while $\mathcal{J}(\br,\bm\Theta)$ is
a \emph{Jastrow factor} which depends on a set of variational parameters $\bm\Theta$.
In $\Psi_{\bm\Theta}(\br)$, the Jastrow factor captures short-range electron
correlation effects, and because the integrals in \Cref{eq:E_expval} are not evaluated
directly, $\mathcal{J}(\br,\bm\Theta)$ admits very complicated functional forms that
typically depend on the inter-electronic distances explicitly.
The mean-field component on the other hand, fixes the nodal structure of the
wavefunction, such that the accuracy of any VMC calculation is ultimately dictated
by $\Phi_0(\br)$, regardless of the choice of the Jastrow factor. A possible way to improve
this situation is to use a \emph{backflow transformation} \cite{LopezRios2006},
which modifies the coordinates of each electron in $\Phi_0(\br)$ based on the position
of all the others
\begin{equation}
  \label{eq:backflow}
  \br_i \to \mathbf{x}_i = \br_i + \bm\xi_i(\br)
\end{equation}
thus moving the position of the nodes.
For electronic problems in real space, the \emph{Slater-Jastrow-backflow} form is
currently the default choice. However, we shall see later on in this chapter how
neural networks can improve upon it.
To obtain the best possible energy $E$ with the ansatz $\Psi_{\bm\Theta}(\br)$,
the expectation value of \Cref{eq:E_expval} is minimized with respect to the
variational parameters $\bm\Theta$.
This is done by starting from an initial set $\bm\Theta^{(0)}$, which is
updated in an iterative fashion. At iteration $t$, the new parameters for
$t+1$ are obtained with
\begin{equation}
  \label{eq:theta_update}
  \bm\Theta^{(t+1)} = \bm\Theta^{(t)} - \mathcal{F}(\bm\Theta^{(t)})
\end{equation}
where the function $\mathcal{F}(\bm\Theta)$ takes different forms depending on
the particular numerical technique chosen.
The simplest option, which is also widely used in the machine learning community,
is gradient descent and its stochastic version.
In this case, the parameters are optimized by taking steps along the direction
pointed by the negative of the energy gradient
\begin{equation}
  \label{eq:GD}
  \bm\Theta^{(t+1)} = \bm\Theta^{(t)} - \alpha\nabla_{\bm\Theta}E(\bm\Theta^{(t)})
\end{equation}
where $\alpha$ is a scalar value determining the step size.
In practice, a first-order scheme such as gradient descent, while computationally
cheap, might require hundreds of iterations to reach convergence.
A more robust option typically used in VMC is the stochastic reconfiguration (SR)
approach by \textcite{Sorella2007}, however its details are beyond the scope of
this introduction, hence they will not be discussed here.

To summarize, a VMC calculation consists in the following steps:
\begin{enumerate}
  \item Obtain $\Phi_0$ and apply the backflow transformation to the
  electronic coordinates
  \item Initialize randomly the Jastrow factor parameters to $\bm\Theta^{(0)}$
  \item Perform a Monte Carlo sweep and sample $N_s$ positions through the
  Metropolis-Hastings algorithm
  \item Compute the energy expectation value according to \Cref{eq:VMC_energy}
  \item Check if the energy is converged (or a predefined maximum of MC sweeps
  is reached)
  \item Terminate the calculation if converged, otherwise continue to the next step
  \item Compute the gradients, update the parameters with \Cref{eq:theta_update}
  and go back to step 3.
\end{enumerate}

The major advantage of VMC over other techniques is that it circumvents the analytical
integration of \Cref{eq:E_expval}, allowing for the very expressive and complicated
wavefunction ansätze based in machine learning models, \textit{e.g.}
neural networks.
At last, we should note that even though we have presented VMC as a real space
approach, the same technique can be used to optimize wavefunctions defined in
any other space, such as that of many-body configurations.

\subsection{Modeling the wavefunction in Fock space}

We have seen in the beginning of this chapter that the first step for the practical
resolution of \Cref{eq:TISE} is the introduction of a finite basis of
many-body configurations, \Cref{eq:FCI_ONV}.
It is not surprising then, that the first major successful attempts to encode the
wavefunction using a ML-inspired approach were obtained in this framework.
In fact, working in Fock space presents several advantages, for instance in the
interpretation of the correlation features and the physics underpinning the
studied systems. After all, this is the basis in which most of the modern
computational quantum physics and chemistry has been developed, such that we
have come a long way in interpreting and analyzing results in these terms.
In the following we shall discuss two unsupervised approaches to the solution
of the Schrödinger equation, one based on a parametric model --- the
neural-network quantum state ansatz --- and one based on a non-parametric one
--- the Gaussian process state ansatz.

\subsubsection{Neural-network quantum state}

The main idea behind the neural-network quantum state ansatz is to use a
neural network to represent the wavefunction. There are many types
of network architectures that can be used for this purpose, however, here we
will focus on the most famous one introduced by \textcite{Carleo2017},
the restricted Boltzmann machine, and only briefly discuss possible extensions
and alternatives.

\begin{figure}
  \centering
  \includegraphics[width=0.65\textwidth]{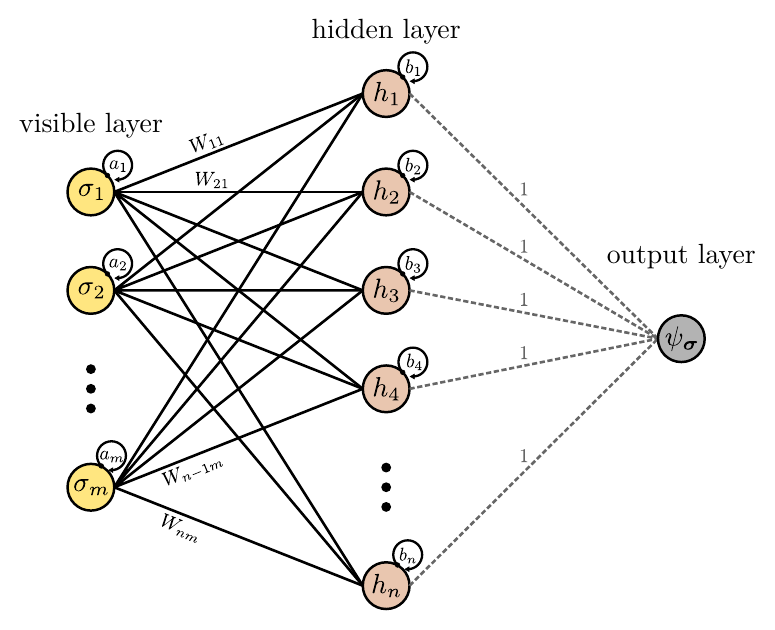}
  \caption{The RBM network architecture underpinning the NQS ansatz
  introduced by \textcite{Carleo2017}. The weights between the hidden
  and outputs layers are fixed to unity, and simply act as a sum.}
  \label{fig:RBM}
\end{figure}

\textit{Restricted Boltzmann machines.}\newline
A restricted Boltzmann machine is a generative model originally developed to
represent classical probability functions, and that can be used to generate
new samples according to the distribution learned from the underlying data.
This is achieved by defining an energy function
\begin{equation}
  \label{eq:E_RBM}
  E(\bv,\bh) = -\sum_{j=1}^{m} a_j v_j -
  \sum_{i=1}^{n} b_i h_i -
  \sum_{i=1}^{n}\sum_{j=1}^{m} h_i W_{ij} v_j
\end{equation}
which depends on interconnected visible and hidden binary variables $v_j,h_i \in \{-1,1\}$,
respectively. Their joint probability density $p(\bv,\bh)$ is assumed to follow
the Boltzmann distribution
\begin{equation}
  \label{eq:p_RBM}
  p(\bv,\bh) = \frac{1}{Z} e^{-E(\bv,\bh)}
\end{equation}
where $Z$ is the partition function defined in the same way as in statistical physics,
ensuring that the probabilities sum up to one.
The crucial breakthrough has been to interpret the marginal distribution over the
visible units as wavefunction amplitudes, that is
\begin{equation}
  \label{eq:pm_RBM}
  p(\bv) = \sum_{\bh} p(\bv,\bh) = \psi(\bv)
\end{equation}
In this context, the vector $\bv$ of visible units represents a many-body configuration
$\bs$, whose correlation is mediated by the hidden variables $\{h_i\}_{i=1}^m$.
This architecture, shown in \Cref{fig:RBM}, results in the following ansatz for the
wavefunction amplitudes (note the switch in notation from $\bv$ to $\bs$ for the
input vector)
\begin{equation}
  \label{eq:RBM_psi_exp}
  \psi(\bs) = \sum_{\bh} e^{\sum_j a_j \sigma_j +
  \sum_i b_i h_i + \sum_{i,j} h_i W_{ij} \sigma_j}
\end{equation}
where, without loss of generality, the normalization factor $Z^{-1}$ from
\Cref{eq:p_RBM} was dropped for simplicity.
Thanks to the fact that there is no \emph{intralayer} connection among the hidden
units, the external sum in \Cref{eq:RBM_psi_exp} can be analytically
traced out, yielding the expression
\begin{equation}
  \label{eq:RBM_psi_cosh}
  \psi(\bs) = \prod_{j=1}^m e^{a_j\sigma_j} +
  \prod_{i=1}^n 2 \cosh \Big( b_i + \sum_{j=1}^m W_{ij} \sigma_j \Big)
\end{equation}
which can be efficiently evaluated for each $\bs$.
\Cref{eq:RBM_psi_exp,eq:RBM_psi_cosh} are manifestly non-negative.
This is a necessary requirement for modeling probability distributions, however,
in the context of fermionic wavefunctions, constitutes a shortcoming.
Indeed, for a complete description of a quantum state, both the amplitude and the
phase factor of the wavefunction are needed, such that in practice, the network
weights are required to admit complex numbers. This is the approach used in the
original work \cite{Carleo2017}, however extensions to more complicated architectures
explicitly encoding the sign of the amplitude \cite{Xia2018,Cai2018} or based on
deep Boltzmann machines are also possible \cite{Carleo2018,Yang2020}, allowing the
weights to remain real numbers.\newline
From a physics standpoint, we can understand the neural network of \Cref{fig:RBM} by
thinking of the particle(s) sitting on the local state $\sigma_j$ as interacting with
the ones described by $\sigma_{k\neq j}$ through the \emph{auxiliary degrees of freedom}
provided by the hidden units $h_i$.
In this framework, the weights $\mathbf{W}$ and bias vectors $\mathbf{a}$ and $\mathbf{b}$
regulate the magnitude of these interactions, effectively encoding correlation
between the particles.
By increasing the number $n$ of hidden units, or equivalently the hidden-unit density
$\alpha = \tfrac{n}{m}$, the wavefunction becomes more expressive and can describe
higher-order correlation features. Importantly, while the number of possible
Fock space configurations increases exponentially with the system size $m$,
the number of parameters defining the RBM only grows as $\mathcal{O}(nm + n + m)$.
The ability to predict the amplitude $\psi(\bs)$ of exponentially many
configurations $\bs$, with a parametrization that only scales polynomially, is one
of the key features underlying the RBM ansatz.
Ultimately, the accuracy is dictated by the density $\alpha$, and in principle,
there is no upper limit to the width of the hidden layer. In fact, RBMs are known
as \emph{universal approximators} \cite{LeRoux2008}, capable to reproduce any
probability distribution to arbitrary accuracy, provided enough hidden units are
included.\newline
It is important to realize that the parametrization achieved through
\Cref{eq:RBM_psi_cosh} does not represent a wavefunction expressed in an actual
many-particle basis. Instead, it just encodes the amplitudes
$\psi_{\bs} = \psi_{RBM}(\bs)$ of the expansion given in \Cref{eq:FCI_ONV}.
Explicitly expressing $\Psi$ in this way completely nullifies the advantages
of the RBM parametrization because of the exponential cost incurred for
evaluating all quantities of interest, \textit{e.g.} expectation values.
For this reason, variational Monte Carlo is used to evaluate the latter and
optimize the parameters.

\textit{Energy evaluation and RBM optimization.}\newline
In typical applications, we are interested in finding the ground state of a
many-body system described by the Hamiltonian $\hat{H}$. In absence of samples
of the exact wavefunction $\Psi$, it is not possible to learn the optimal
parameters $\bm\Theta = (\mathbf{W},\mathbf{a},\mathbf{b})$ in a supervised fashion,
instead, we rely on a reinforcement learning algorithm based on the variational
principle.
Here, we essentially follow the variational Monte Carlo strategy outlined in
the previous section, albeit in Fock space rather than in real space.
\newline
The parameters of the network are initialized with random values and then
a Markov chain of many-body configurations,
\begin{equation}
  \label{eq:NQS_MCMC}
  \bs^{(1)} \to \bs^{(2)} \to \ldots \to \bs^{(N_{s})}
\end{equation}
is generated according to the Metropolis-Hastings algorithm. That is,
at each step $k$ of the random walk, a new configuration $\bs^{(k+1)}$ is
generated by randomly changing the state of a random local degrees of
freedom $\sigma_i$ of the current configuration $\bs^{(k)}$.
For example, a spin is flipped at site $i$, going from $\sigma_i=\uparrow$
to $\sigma_i=\downarrow$.
This step is then accepted with probability
\begin{equation}
  \label{eq:NQS_MH}
  A(\bs^{(k)}\to\bs^{(k+1)}) = \min
  \bigg(1,\left| \frac{\Psi_{RBM}(\bs^{(k+1)})}{\Psi_{RBM}(\bs^{(k)})} \right|^2\bigg)
\end{equation}
meaning that if the new configuration $\bs^{(k+1)}$ with the flipped spin
has a larger amplitude than $\bs^{(k)}$, it is accepted with 100\% probability,
otherwise with a probability proportional to their ratio.
This approach is known as \emph{Markov chain Monte Carlo} (MCMC).
After a Monte Carlo sweep ($N_s$ accepted steps), the energy $E$ is evaluated
according to \Cref{eq:VMC_energy}, with many-body configurations $\bs^{(k)}$ instead
of electronic coordinates $\br^{(k)}$.
At the same time, in a completely analogous manner, the gradient of the energy
$\nabla_{\bm\Theta}E$ is stochastically sampled.
This allows to obtain a new set of parameters according to either gradient descent
or stochastic reconfiguration.
This procedure is repeated for $N_{MC}$ sweeps or until the energy and gradients
do not change significantly anymore.

One of the advantages of the RBM architecture is the computational complexity
associated with the evaluation of the energy and gradients.
The cost to compute the wavefunction amplitude of a many-body configuration scales as
$\mathcal{O}(n+m)$ if the effective angles $\theta_i=b_i+\sum_{j=1}^m W_{ij}\sigma_j$
are computed all at once at the beginning and kept in memory for the entire
procedure.
This means that the evaluation of \Cref{eq:NQS_MH} has the same asymptotic cost
of $\mathcal{O}(n+m)$.
This process is repeated $N_s$ times for each MC sweep, whereby the effective
angles are updated one by one after each accepted step, with a constant cost of
$\mathcal{O}(1)$. This totals to $\mathcal{O}(N_s(m+n))$ for each sweep.
The evaluation of the local energies and gradients carries the same computational
cost as the evaluation of the amplitudes.
Depending on the choice of optimization algorithm, the scaling may be linear in the
number of variational degrees of freedom for first-order methods such as
gradient descent, or quadratic for stochastic reconfiguration.
For the latter, the most demanding step is the solution of the linear system of
equations to invert the covariance matrix, which
scales as $\mathcal{O}((mn+m+n)^2 N_s)$. However, this can be reduced to linear
in $N_{var}=mn+m+n$ by exploiting the product structure of the covariance matrix.
Overall, considering that the calculation entails $N_{MC}$ sweeps, the complexity
of the NQS ansatz based on RBMs scales as $\mathcal{O}((mn+m+n)N_s N_{MC})$.

\textit{Physics applications \& properties of the NQS ansatz.}\newline
The NQS ansatz based on restricted Boltzmann machines and extensions thereof has
been applied to a variety of model systems with great success. State-of-the-art
accuracy was reached for several spin Hamiltonians, such as the transverse field
Ising chain, the antiferromagnetic Heisenberg model,
as well as both the bosonic and fermionic Hubbard models
\cite{Carleo2017,Nomura2017,Carleo2018,Cai2018,Melko2019,Vieijra2020,Choo2020}.
For RBMs, systematic convergence to the exact results can be achieved by
increasing the hidden-unit density $\alpha$.
Impressively, in the one-dimensional Heisenberg chain, the RBM surpasses the
accuracy of other state-of-the-art approaches, such as DMRG, with approximately
three orders of magnitude fewer parameters \cite{Carleo2017}, highlighting the
high degree of compression achievable by this representation.
Symmetries can be included in a straightforward manner, by
summing over all symmetry operations $\mathcal{S}$ that the ansatz has to respect,
that is
\begin{equation}
  \label{eq:RBM_psi_cosh_sym}
  \tilde{\psi}(\bs) = \sum_{\mathcal{S}} \psi_{RBM}(\mathcal{S}\bs)
\end{equation}
In practice, \Cref{eq:RBM_psi_cosh_sym} can be recast into a RBM with a hidden
layer of $m \times S$ units, where $S$ is the total number of symmetry operations.
Abelian and non-abelian symmetries have been implemented, providing access to
excited states, a better overall accuracy at fixed hidden-unit density $\alpha$
compared to the non-symmetric RBM architectures, and better convergence properties
thanks to the reduced size of the variational parameters space
\cite{Carleo2017,Choo2018,Vieijra2020,Nomura2021}.\\
Of particular interest is the extension of RBMs to architectures that include
more than one layer. Deep Boltzmann machines (DBMs) have been shown to
be more general than RBMs, thereby exactly representing certain quantum
mechanical states in a compact form, which would otherwise not be possible
with RBMs \cite{Gao2017,Carleo2018}.
For instance, the increased flexibility of DBMs allows to encode the phase
of the wavefunction avoiding the use of complex algebra, even though completely
separating phase and amplitudes is a viable option too \cite{Szabo2020}.
On the other hand, the presence of more than one layer does not allow
to trace out the hidden degrees of freedom as done from \Cref{eq:RBM_psi_exp}
to \Cref{eq:RBM_psi_cosh}, thus increasing the computational cost of the
forward pass to evaluate the neural network output.

\textit{Quantum chemical applications.}\newline
The structure shown in \Cref{fig:RBM} is reminiscent of low-rank tensor network
states \cite{Orus2019}, however, an important feature that sets RBMs apart from
the latter is the intrinsic non-local nature of the connections induced by the
hidden units. This allows to describe systems of arbitrary dimensions and physics
containing long-range interactions, \textit{e.g.} molecules governed by the full
\textit{ab initio} Hamiltonian, central to the electronic structure problem in
quantum chemistry.
Starting from the second-quantized form of the many-body fermionic Hamiltonian
\begin{equation}
  \label{eq:NQS_H_sq}
  \hat{H} = \sum_{pq} t_{pq} \hat{a}_p^\dagger \hat{a}_q +
  \sum_{pqrs} V_{pqrs} \hat{a}_p^\dagger\hat{a}_r^\dagger\hat{a}_s\hat{a}_q
\end{equation}
the problem can be recast into a spin basis by a Jordan-Wigner
transformation \cite{Jordan1928}. In \Cref{eq:NQS_H_sq}, $t_{pq}$ and
$V_{pqrs}$ are one-electron and antisymmetrized two-electron integrals,
respectively, while $\hat{a}_p^\dagger$ ($\hat{a}_q$) is the creation (annihilation)
fermionic operator for spin-orbital $\phi_p$ ($\phi_q$).
The resulting transformed Hamiltonian (see \textcite{Xia2018,Choo2020} for more
details) can then be studied with the NQS ansatz without further modifications
and following the same strategy outlined in the previous subsection.\newline
In complete analogy to traditional quantum chemical methods, this approach relies
on the introduction of a finite one-particle basis set, with the NQS ansatz providing
a compact representation of the full CI wavefunction.
The first example applications were on the dissociation of small diatomic systems,
such as \ce{H_2}, \ce{C_2}, \ce{N_2} and \ce{LiH}, or on the ground state optimization
of \ce{NH_3} and \ce{H_2O} at the equilibrium geometry \cite{Xia2018,Choo2020}.
In all cases, in combination with the minimal STO-3G basis set, the accuracy
was on-par or superior to CCSD and CCSD(T) at all correlation regimes, highlighting
the flexibility of this approach.
The ground state energies at the equilibrium geometries for all
these systems are shown in \Cref{tab:RBM_results}.
\begin{table}
  \centering
  \caption{Ground state energy differences with respect to full CI
  for several molecules at the equilibrium geometry.
  The energies are obtained in combination with a STO-3G basis set
  and are given in Hartree.
  The hidden-unit density for the first four molecules was $\alpha = 1$,
  while for the last two, $\alpha = 2$.
  Data taken from \textcite{Choo2020}.}
  \begin{tabular}{lccc}
  \toprule
  Molecule & CCSD   & CCSD(T) & RBM \\
  \midrule
  H2       & 0.0    & 0.0     & 0.0 \\
  LiH      & 0.0000 & 0.0     & 0.0002 \\
  NH3      & 0.0002 & 0.0001  & 0.0005 \\
  H2O      & 0.0002 & 0.0001  & 0.0001 \\
  C2       & 0.0163 & 0.0032  & 0.0016 \\
  N2       & 0.0057 & 0.0036  & 0.0007 \\
  \bottomrule
  \end{tabular}
  \label{tab:RBM_results}
\end{table}
Importantly, for all systems but \ce{C_2} and \ce{N_2}, an RBM with a
hidden-unit density $\alpha = 1$ was used, with energies differing by less than a
milliHartree from the coupled cluster values.
By increasing $\alpha$ to 2, the RBM outperforms the coupled cluster approaches,
suggesting that the same would likely happen for the smaller systems as well.\\
Owing to the sharply peaked distribution underlying the wavefunction configurations,
that is, the weights associated to the Slater determinants decrease very quickly
in magnitude,
the Markov chain Monte Carlo sampling with the Metropolis-Hastings algorithm is much
less effective for larger basis sets, with extremely low numbers of accepted steps.
This remains an open problem, and different sampling strategies needs to be devised
for studying larger systems \cite{Choo2020}. Recent efforts in this direction involve
a new class of generative models called autoregressive neural
networks \cite{Wu2021,Barrett2021}.
Restricted Boltzmann machines and deep Boltzmann machines with two and three hidden
layers were also used as the active space (AS) solver in CAS-CI calculations
\cite{Yang2020}.
In this case, the visible layer units are mapped directly to the occupation number
of the spin orbitals, and two neural networks are optimized separately; one for the
phase of the wavefunction and one for the amplitudes. In this way, the weights for
the amplitudes do not have to be complex numbers to encode the phase, as discussed previously.
This method was tested on the indocyanine green molecule, for active spaces of
increasing size, from four electrons in four orbitals to eight electrons in eight
orbitals, and on the dissociation of \ce{N_2} with an AS of six electrons in six
orbitals.
The energies obtained were in general within 10 microHartree from the deterministic
full CI solver, and the accuracy was more consistent for the deep models rather
than the shallow one-layer RBM, highlighting also in the quantum chemical context
the greater representational power provided by DBMs over RBMs.

These few examples of quantum chemical applications highlight the potential of
neural-network quantum states in electronic structure theory. While these are first
proof-of-concept works, they already provide important information on the issues
that need to be addressed for a more tight integration with traditional methods.
The use of a single-particle basis set provides the advantage of an implicitly
correct antisymmetric wavefunction, and an easy integration with existing quantum chemical
packages.
On the other hand, the same limitations apply too. While RBMs provide very compact
representations, the exponential scaling of the Fock space will eventually
limit the size of the systems to which the NQS ansatz can be applied to.
Their use as active space solvers is likely to be their best scope of application,
in the same way as the MPS wavefunction has been so far within electronic structure
theory. In this comparison, it seems that
RBMs have some representational advantages over the latter, thanks to their long
range connections between the units. On the other hand, the use of a stochastic
sampling algorithm might be a limitation, in particular for larger bases and
systems mostly dominated by a few many-body configurations.

\subsubsection{Gaussian process state}

The success achieved in tackling the quantum many-body problem by parametric models
based on neural networks, naturally makes one wonder if non-parametric
ones can be equally effective.
Such an example is the Gaussian process state ansatz, a wavefunction representation
based on Gaussian process (GP) regression \cite{Glielmo2020,Rath2020}.
GP regression falls under the umbrella of kernel methods, discussed in
chapter 9, but also within the framework of Bayesian inference, presented in
chapter 10.
In this book, we have already encountered several examples of Gaussian processes in
action, for instance in the context of machine learning potentials in
chapter 13, which can be used for classical molecular dynamics, or as a
surrogate model for geometry optimization and transition state search in
chapter 17.
Here, on the other hand, it is used to model the wavefunction in Fock space.
In GP regression, the functional form of the model is constrained by the choice
of the kernel, which defines the basis functions of the linear expansion, and
the size of the training set, which determines the total number of parameters
available for optimization.
This \emph{data-driven} feature has the great advantage that for an increasing amount
of data, the model systematically converges to the exact generating function.
In contrast, parametric methods such as neural networks are constrained from the
outset by the initial choice of parameters, \textit{e.g.} the hidden-unit density
$\alpha$ in RBMs, and thus, even in the limit of an infinite training set, might
not be able to exactly reproduce the data.
The GPS ansatz was developed in the framework of lattice systems, in particular
focusing on the Fermi-Hubbard Hamiltonian. Hence, in this subsection we will restrict
our attention to this type of application, only briefly discussing the potential
extension to treat molecular systems.

\begin{figure}
  \centering
  \includegraphics[width=10cm]{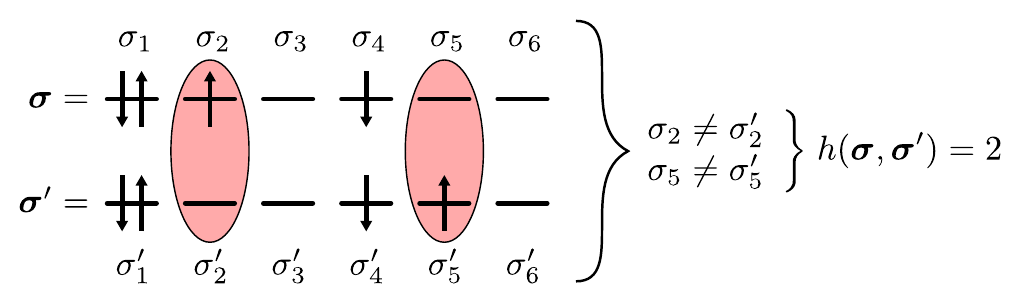}
  \caption{Hamming distance between two different many-body configurations
  of a 6-site 1D chain. The configurations are different at two sites, hence
  their Hamming distance is 2.}
  \label{fig:Hamming}
\end{figure}

\begin{figure}
  \centering
  \includegraphics[width=8cm]{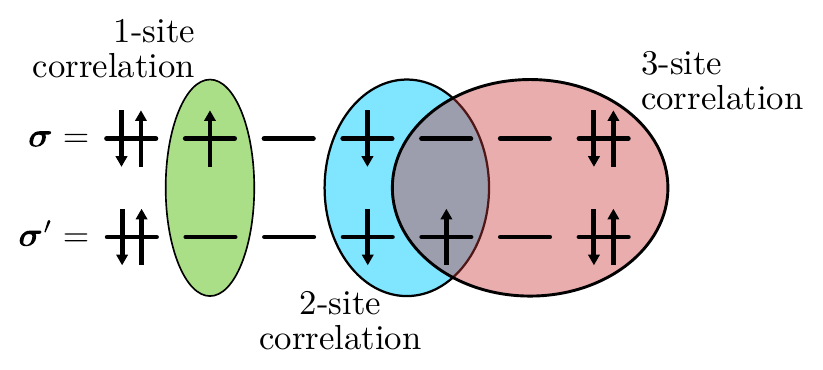}
  \caption{Correlation features identified by the kernel. Note that $n$-site
  correlation features with $n>1$ are not only compared between groups of
  $n$ neighboring sites, but with all possible combinations of $n$ sites.}
  \label{fig:kernel_features}
\end{figure}

\textit{The ansatz.}\newline
A Gaussian process state models the wavefunction amplitudes of a quantum state
as an exponential of a Gaussian process, that is
\begin{equation}
  \label{eq:GPS_Psi}
  \Psi_{GPS}(\bs) = e^{\sum_{b=1}^{N_b} k(\bs,\bs_b') w_b}
\end{equation}
where the function $k(\bs,\bs')$ is the \emph{kernel} and the elements $w_b$ form
a weight vector $\bw$ of variational degrees of freedom.
The set $\{\bs_b\}_{b=1}^{N_b}$ constitutes the \emph{data} underpinning the
GP regression model, and can be understood as a basis set of reference configurations.
The defining idea of the GPS ansatz is to encode the complicated many-body
correlation effects among the particles through the kernel function.
This happens \textit{via} a scalar product between $\bs$ and $\bs'$, embedded in
a high-dimensional feature space dictated by the choice of $k(\bs,\bs')$.
Clearly, the form of the kernel function and the set of reference configurations
$\bs_b$ are crucial ingredients prescribing the accuracy of this ansatz.
After all, \Cref{eq:GPS_Psi} reveals that $\ln(\Psi_{GPS}(\bs))$ is just a linear
expansion in the basis spanned by the kernel functions centered at the reference
configurations $\{\bs_b\}_{b=1}^{N_b}$.
A kernel function that provides sufficient flexibility to model the correlation
between the particles to any order is given by
\begin{equation}
  \label{eq:GPS_kernel}
  k(\bs,\bs') = \exp\left(\frac{-h(\bs,\bs')}{\theta|r_i-r_0|^\gamma}\right)
\end{equation}
where $h(\bs,\bs')$ is the \emph{Hamming distance} between two many-body
configurations, $r_i$ and $r_0$ denote positions on the lattice and $\theta$
and $\gamma$ are two \emph{hyperparameters}.
The Hamming distance is defined as
\begin{equation}
  \label{eq:GPS_Hamming}
  h(\bs,\bs') = \sum_{i=1}^m (1 - \delta_{\sigma_i \sigma_i'})
\end{equation}
and quantifies the similarity between two many-body configurations by comparing
their local occupations. For every site $i$ where the local states
$\sigma_i$ and $\sigma_i'$ differ, the distance increases by one.
An example is shown in \Cref{fig:Hamming}.
The absolute value in the denominator, $|r_i-r_0|^\gamma$, measures how far apart
on the lattice is site $i$ (at position $r_i$) from a reference site $0$
(at position $r_0$) that is chosen arbitrarily. For a positive value of $\gamma$, this
expression suppresses differences between sites that are distant from the reference,
favoring short-range correlations.
For translationally invariant systems, the choice of the reference site is
irrelevant, while for systems without symmetry, an additional sum over all possible
reference sites should be added in \Cref{eq:GPS_kernel}.
The second hyperparameter, $\theta$, is best understood by taking the Taylor
expansion of \Cref{eq:GPS_kernel}, that is
\begin{equation}
  \label{eq:GPS_k_taylor}
  k(\bs,\bs') = 1 - \underbrace{\frac{\sum_i 1 -
  \delta_{\sigma_i,\sigma_i'}}{\theta}}_{\text{1-site correlation}} +
  \underbrace{\frac{\sum_{i,j} (1 - \delta_{\sigma_i,\sigma_i'})
  (1 - \delta_{\sigma_j,\sigma_j'})}{2\theta^2}}_{\text{2-site correlation}}
  + \ldots
\end{equation}
where we omitted from the expansion the term $|r_i-r_0|^\gamma$ for clarity.
\Cref{eq:GPS_k_taylor} shows how the kernel computes \emph{correlation features}
of increasing order (or rank) between the configurations.
The first non-trivial term compares sites one by one, hence extracting 1-site
correlation features. The second term compares sites in pairs, hence capturing
2-site correlation features, and so forth. This order-by-order comparison
is graphically exemplified in \Cref{fig:kernel_features}.
With this understanding of the kernel, the denominator of \Cref{eq:GPS_k_taylor}
unfolds the role of $\theta$ in suppressing higher-rank correlation features
for values larger than 1. On the other hand, the opposite happens for $\theta<1$.
The expansion in \Cref{eq:GPS_k_taylor} also reveals a key feature of the kernel
function. Truncation of the sum to a fixed order allows the GPS to reproduce other
known wavefunction ansätze. For instance, a first-order truncation corresponds to
a Gutzwiller-type wavefunction \cite{Gutzwiller1963}, where only single-site
occupancies are compared. Keeping only the second-order term, a generalized Jastrow
representation is recovered \cite{Jastrow1955}, correlating all possible site pairs.
Other, more general ansätze can also be expressed through the GPS by restricting
site comparisons within a maximum range from the reference site. This generates,
for instance, entangled plaquette states \cite{Mezzacapo2009} and correlator product
states \cite{Changlani2009}.
All these traditional approaches explicitly parametrize the wavefunction in a
particular feature space by constraining the rank or the range of the correlations
modeled (note that this is analogous to the excitations classes of quantum chemical
methods such as CCSD, where the singles would correspond to 1-site correlations, the
doubles to 2-site correlations, and so forth). In strong contrast, the GPS implicitly
extracts these features from the full kernel through the parameters associated to the
reference configurations $\{\bs_b\}_{b=1}^{N_b}$.
Crucially, while the number of multi-site features increases exponentially with
the number of sites, the evaluation of \Cref{eq:GPS_kernel} only scales polynomially,
highlighting the advantage of this kernel-based approach over explicit parametrizations.
Furthermore, such a representation always ensures that the exact wavefunction can be
obtained in the limit of the complete set of reference configurations (the
complete set would correspond to a full CI).

\begin{figure}
  \centering
  \includegraphics[width=.7\textwidth]{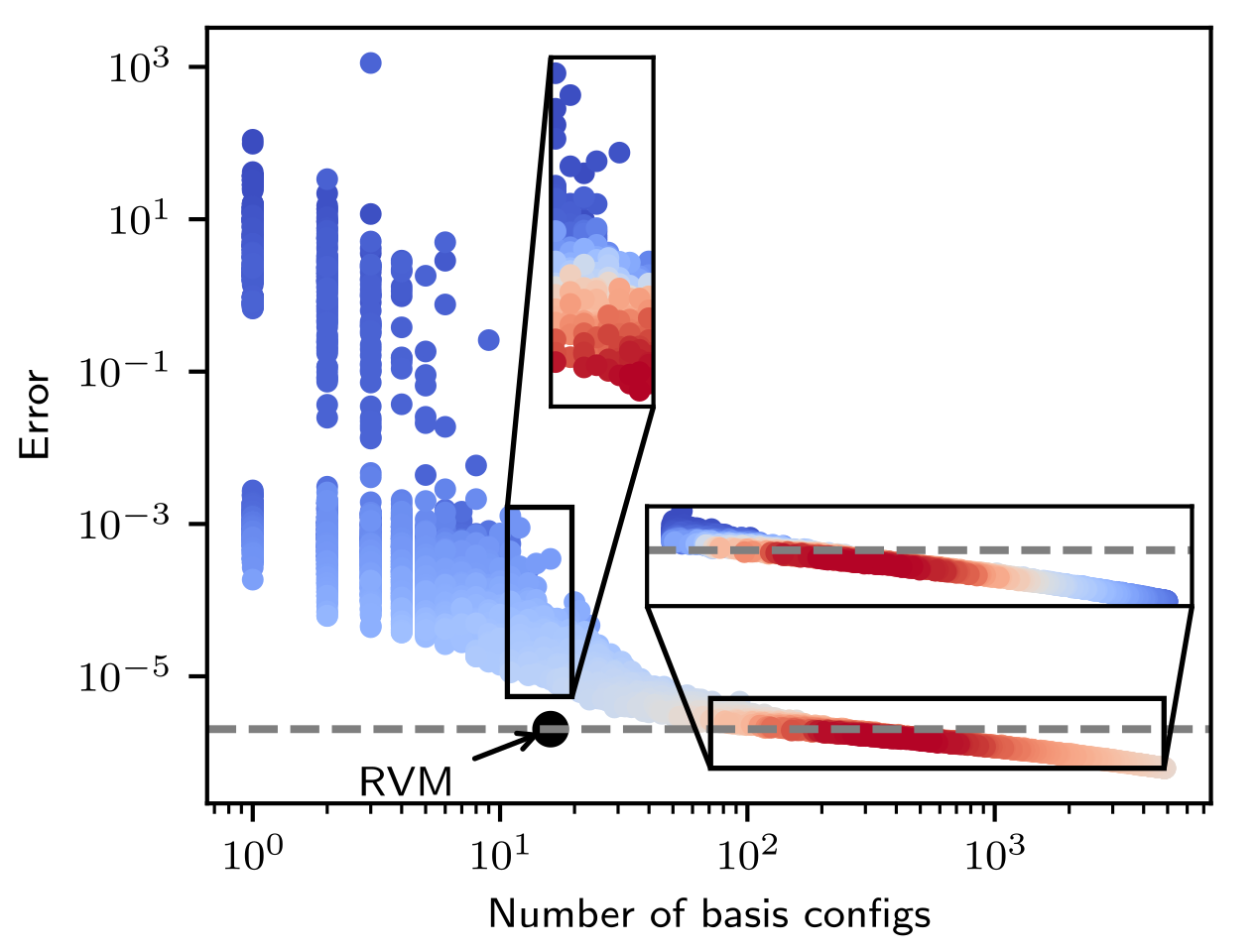}
  \caption{Mean squared error between
  the GPS ansatz and the exact ground state of the Fermi-Hubbard model
  on an 8-site lattice. The hyperparameters $\gamma$, $\theta$ and
  $\sigma_t$ have been fixed \textit{a priori}.
  Each point represents a random selection of $N_b$ basis configurations
  and the associated error.
  The color of the points represents the value of the log marginal
  likelihood for the weights, with red denoting larger values and blue
  smaller values. The dashed line and the black point show the accuracy
  reached by the RVM algorithm. The color-map is rescaled in the insets.
  Figure taken from \citetitle{Rath2020} \fullcite{Rath2020}
  (DOI: 10.1063/5.0024570).}
  \label{fig:compression_error}
\end{figure}

\textit{Representational power of GPS.}\newline
Before discussing the optimization of the GPS in a variational framework,
it is instructive to investigate the representational power of this ansatz by
approximating an exact wavefunction in a supervised fashion.
In particular, for a fixed choice of basis configurations $\{\bs_b\}_{b=1}^{N_b}$
and hyperparameters $\theta$ and $\gamma$, the GPS ansatz can be trained within a
Bayesian inference framework with a set of training configurations and associated
wavefunction amplitudes $\{\bs_t,\psi(\bs_t)\}_{t=1}^{N_t}$.
In practice, the actual GP is the logarithm of \Cref{eq:GPS_Psi}, that is
\begin{equation}
  \label{eq:GPS_log_psi}
  \varphi_{GPS}(\bs) = \ln(\psi_{GPS}(\bs)) = \sum_{b=1}^{N_b} w_b k(\bs,\bs_b')
\end{equation}
and this model is trained on the log-amplitudes, $\varphi(\bs_t) = \ln(\psi(\bs_t))$,
rather than directly on the wavefunction amplitudes.
The optimal parameters $w_b$ are then obtained as the mean of the posterior
distribution for the weights according to Bayes theorem. This corresponds to a
direct minimization of the squared error between the exact log-amplitudes and the
predicted ones
\begin{equation}
  \label{eq:GPS_Bayes}
  \bw^* = \argmin_{\bw} \sum_{t=1}^{N_t}
  \frac{|\varphi(\bs_t)- \varphi_{GPS}(\bs_t)|^2}{\sigma_t^2}
\end{equation}
where $\sigma_t^2$ is a variance hyperparameter that regulates how tightly the
GP reproduces the exact log-amplitudes at the training points (see
\textcite{Rath2020} for a more detailed discussion).
With such a scheme, it is possible to evaluate how accurately the GPS can represent
the wavefunction of a known quantum state as a function of the number of basis
configurations.
For example, the plot in \Cref{fig:compression_error} shows the mean squared error
of the GPS with respect to the exact ground state of the 8-site Fermi-Hubbard model
in the strong correlation regime ($U/t=8$).
As can be seen, the choice of basis configurations plays a crucial role for the
accuracy of the GPS ansatz. In particular, for a basis set containing a few
many-body configurations, the error can vary across several orders of magnitude,
whereas for large basis sets the accuracy of the wavefunction is much more
consistent.
This can be understood in terms of the implicit parametrization of the wavefunction
through the data $\{\bs_b\}_{b=1}^{N_b}$; a representative set of many-body
configurations is needed for modeling the most important correlation features.
It is thus clear from \Cref{fig:compression_error} that the choice of the basis
is paramount to obtain a compact and accurate representation of the wavefunction.
A way to identify an optimal basis set $\{\bs_b\}_{b=1}^{N_b}$ is provided by
relevance vector machine (RVM), a sparsification algorithm that selects the
configurations based on the log marginal likelihood for the weights.
Large values of the latter are associated to important configurations, whereas
the opposite is true for small values. The RVM can thus be used to select the
most representative $\bs_b$'s out of a set of candidate configurations, yielding
an optimally compact basis set.
An example of this can be seen in \Cref{fig:compression_error}, where the RVM
algorithm produces a GPS that is significantly more accurate than the random
selection, for a fixed number of basis configurations.
Whereas the RVM algorithm is used to select the basis set, the optimization of
$\theta$ and $\gamma$ can be done through a sampling of the hyperparameters
space via a Bayesian scheme. In this optimization procedure, an initial assumed
distribution is used to sample the hyperparameters, and gets refined after each
iteration with the additional information obtained from the sampled points.

\begin{figure}
  \centering
  \includegraphics[width=8cm]{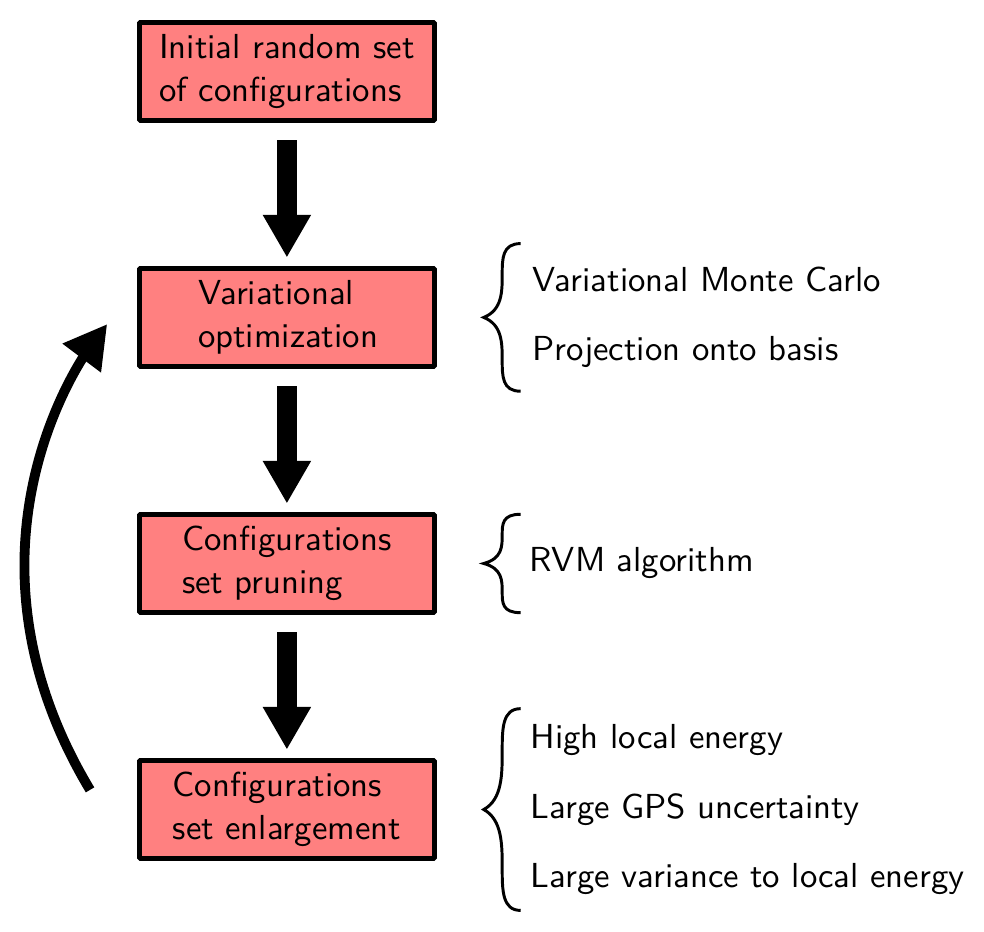}
  \caption{Bootstrap optimization steps of the GPS ansatz with the
  possible algorithmic choices on the right-hand side.}
  \label{fig:GPS_bootstrap}
\end{figure}

\textit{Bootstrapped Optimization.}\newline
After establishing that the GPS is able to compactly represent complicated
wavefunctions, we shall turn to the case where the target quantum state is
not known \textit{a priori}.
In this case, the parameters $w_b$ can be optimized by minimizing the energy
expectation value according to the variational principle.
In absence of data, an initial random set of basis configurations
$\{\bs_b^{(0)}\}$ underpinning the GPS model is selected, and the associated
weights $w_b$ are initialized to zero.
Alternatively, a mean-field or another approximate wavefunction can be used
to pre-train the weights of the GPS, providing a better starting point.
Then, the values of the weights are optimized by minimizing the energy of
the system.
Similar to neural-network quantum states, the GPS ansatz models the
wavefunction amplitudes of a quantum state, such that to evaluate the energy
expectation value, the GPS needs to be projected on a particular basis or
sampled according to a stochastic technique such as variational Monte Carlo.
Next, the reference set of configurations $\{\bs_b^{(0)}\}$ is pruned
through the RVM sparsification procedure, reducing the size of the basis
while keeping the accuracy.
New many-body configurations that are different from the initial ones
then added to the basis, yielding a new reference set $\{\bs_b^{(1)}\}$.
Here, there are various criteria that can be used to enlarge the basis set,
for instance one could select configurations associated to a large uncertainty
of the GPS (obtained through the Bayesian framework, see chapter 10),
to a high local energy, or to a large variance contribution to the local energy.
Regardless of the data augmentation criterion used, the weights of the new
enlarged set are then optimized again variationally, and this process is
repeated until convergence.
A scheme depicting the optimization steps can be seen in \Cref{fig:GPS_bootstrap}.
The optimization procedure just outlined was used to obtain the ground state
of the half-filled Fermi-Hubbard model in one and two dimensions in the strong
correlation regime with $U=8t$ \cite{Glielmo2020}.
For the one-dimensional chain with 32 sites, the GPS ansatz reached a relative
energy error (with respect to DMRG) of $1.2 \times 10^{-3}$ Hartree with
1369 reference many-body configurations.
On a $6\times 6$ square lattice, the GPS systematically converges toward the
exact ground state energy with increasing number $N_b$ of basis configurations.
Both standard approaches based on the Gutzwiller wavefunction and Jastrow with
pair correlations are easily surpassed in accuracy with less than 100
basis configurations, and the results obtained are on-par with the NQS ansatz
(based on the RBM architecture) using a similar number of variational parameters.
Instead, for the larger $8\times 8$ square lattice, the GPS initialized with
a Jastrow wavefunction outperforms the RBM both in accuracy and convergence
rate from the outset. In this case the GPS (with a number of parameters
ranging from 1000 and 2000, depending on the RVM procedure) consistently shows
a lower energy per site throughout the optimization procedure compared to
the Gutzwiller (1 parameter), the Jastrow (34 parameters) and the RBM
(2064 parameters, $\alpha = 16$) ansätze.

\textit{Towards quantum chemical applications of GPS}\newline
The GPS ansatz has been initially used to obtain the ground state of
the Fermi-Hubbard model, however, extensions to tackle the \textit{ab initio}
electronic Hamiltonian are in principle possible.
For instance, following the approach of \textcite{Yang2020} used for NQS
wavefunctions, spin orbital occupation numbers could be used to define the
local Hilbert spaces of the many-body configurations.
While this does not impose any particular restriction on the type of
interactions captured by the RBM (thanks to the non-local nature of the
inter-layer connections), extra attention is required for the GPS model.
In particular, there are several ways to account for the dependence of the kernel
on the distance between the sites, see the denominator of \Cref{eq:GPS_kernel}.
The spin orbitals could simply be mapped onto a one-dimensional lattice like
a matrix product state wavefunction, however this would artificially bias the
range of interactions captured by the kernel in an arbitrary way. On the other
hand, if a basis of localized
orbitals is used, the Euclidean distance between their centers in real space
could be used directly in \Cref{eq:GPS_kernel}.
Another issue is the choice of reference site. Whereas all the sites in a regular
lattice with periodic boundary conditions are equal, the same is not true for the
single-particle basis used for molecules.

Despite these technical details, the GPS ansatz has shown to be a very promising
alternative to other novel ML-based wavefunction approaches such as NQSs.
Above all, the data-driven property intrinsic to non-parametric approaches is
extremely appealing. Instead of constantly modifying the parametrization of
the wavefunction (\textit{e.g.} increasing the hidden-unit density of an RBM)
and rerunning a calculation to reach a desired accuracy, the GPS ansatz can be
systematically refined by the simple addition of new basis configurations and
an extra iteration step, picking up increasingly complex many-body
correlation features.
Compared to traditional quantum chemical methods, the same discussion carried
out for the NQS ansatz applies here as well.
For instance, the limitations intrinsic to one-particle basis sets are present
in this case, which is the price to pay for a method defined in the
second quantization framework. On the other hand, the antisymmetry requirement is
intrinsically satisfied by the many-body configurations basis.
In order to encode the phase of the wavefunction, complex weights can be used
as it was the case for RBMs. To avoid the use of complex algebra, alternative
strategies to encode the phase need to be developed.
At last, it can be envisioned that the GPS will probably find application as
an active space solver in multireference settings, or for accurate benchmark
calculations beyond what is currently feasible with full CI and similar
approaches.

\subsection{Modeling the wavefunction in real space}

In the previous subsection we have seen how parametric and non-parametric
machine learning methods have been used to represent the wavefunction
in Fock space.
A formalism based on second quantization has the great advantage that the
Fermi-Dirac statistics is incorporated by construction in the many-particle
basis, \textit{e.g.} by Slater determinants for typical quantum chemical
approaches. The same is not true for first-quantized methodologies.
In fact, a major obstacle for modeling fermionic wavefunctions directly in
real space is the proper integration of the antisymmetry property in the ansatz.
In this subsection we shall see two examples of neural network architectures
that have succeeded in this respect, with very promising results.

\subsubsection{FermiNet}

The indistinguishability of fermions manifests itself in a wavefunction that
must be antisymmetric with respect to the exchange of two particles, as shown
in \Cref{eq:antisymmetry}.
Traditional quantum chemistry methods incorporate this property by introducing
a basis of many-particle functions that satisfy this requirement, namely Slater
determinants. These have the following general form
\begin{equation}
  \label{eq:FN_SD}
  \Phi(\br) = \mathcal{\hat{A}} \prod_{i=1}^N \phi_i(\br_i) =
  \begin{vmatrix}
    \phi_1(\br_1) & \ldots & \phi_N(\br_1) \\
    \vdots          &        & \vdots          \\
    \phi_1(\br_N) & \ldots & \phi_N(\br_N)
  \end{vmatrix}
\end{equation}
where $\phi_i(\br_j)$ are one-particle functions (spin orbitals) and
$\mathcal{\hat{A}}$ is an \emph{antisymmetrizer} that sums over all possible
pairwise permutations of the particles, multiplied by either $+1$ or $-1$
depending on the parity of the permutation.
This construct automatically satisfies \Cref{eq:antisymmetry}, and conveniently
encodes the exchange of two particles, say $\br_i$ and $\br_j$, by swapping
rows $i$ and $j$ in the determinant.
It is this exactly property that stands at the heart of FermiNet.

\begin{figure}
  \centering
  \includegraphics[width=\textwidth]{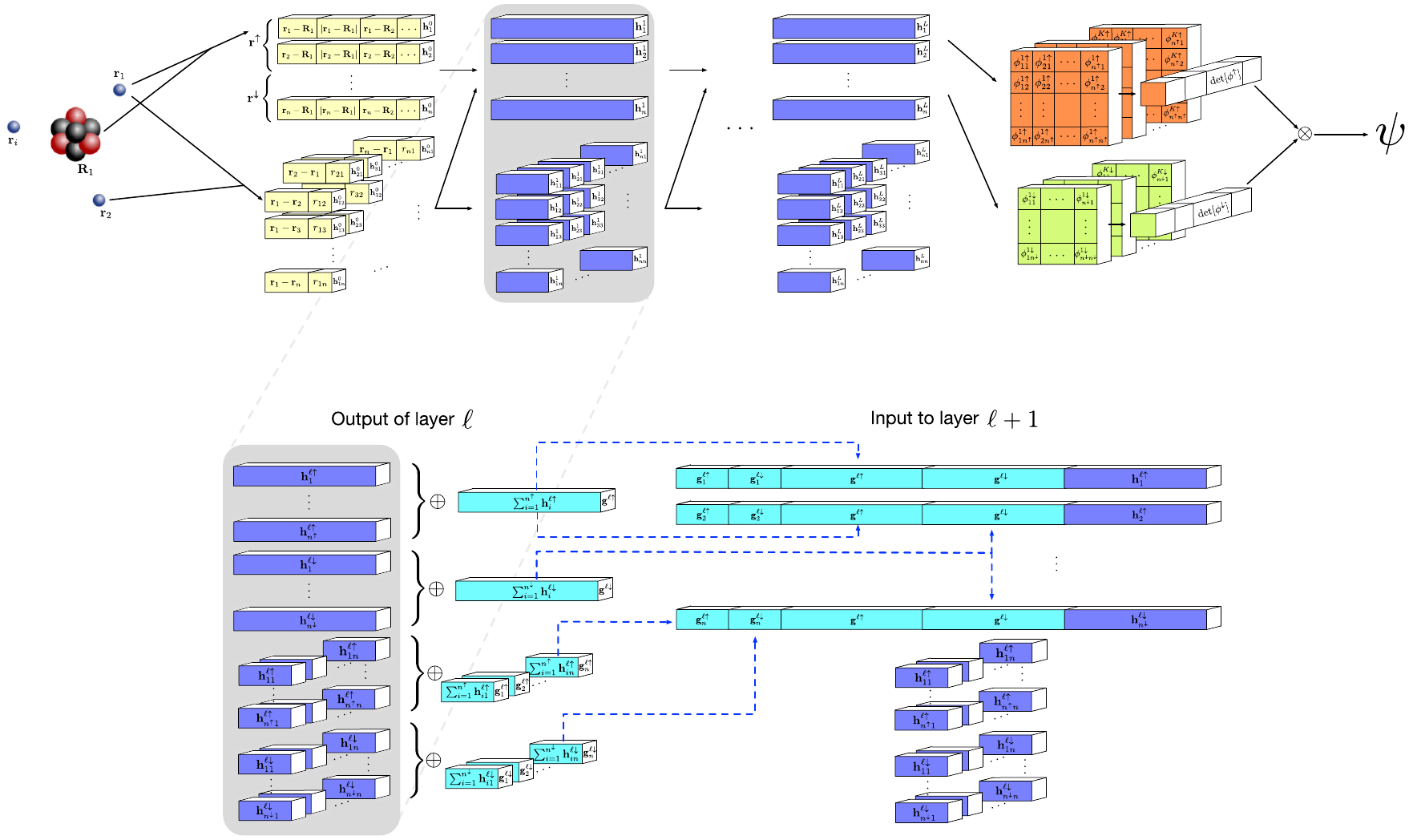}
  \caption{The neural network architecture underlying the FermiNet ansatz.
  Image adapted from \citetitle{Pfau2020} \fullcite{Pfau2020}
  (DOI: 10.1103/PhysRevResearch.2.033429).}
  \label{fig:FermiNet}
\end{figure}

\textit{The ansatz.}\newline
FermiNet is a wavefunction ansatz based on a deep neural network architecture
\cite{Pfau2020}.
The main idea underlying it comes from the realization that the basis functions
$\phi_i(\br_j)$ in \Cref{eq:FN_SD} do not necessarily have to be single-particle
orbitals. The only requirement is that the wavefunction changes sign upon
exchange of two rows in the determinant.
With this observation, the orbitals $\phi_i(\br_j)$ can actually be replaced
by many-electron functions of the form
\begin{equation}
  \label{eq:FN_PEMO}
  \phi_i(\br_j;\br_{/j}) =
  \phi_i(\br_j;\br_1,\ldots,\br_{j-1},\br_{j+1},\ldots,\br_N)
\end{equation}
with the property that swapping the positions $\br_k$ and $\br_l$ of any two
particles with $k \neq l \neq j$, leaves the sign of $\phi_i(\br_j;\br_{/j})$
unchanged.
Slater determinants constructed with these \emph{permutation-equivariant} functions,
$\tilde{\Phi}(\br) = \mathcal{\hat{A}}\prod_{i=1}^N \phi_i(\br_i;/\br_i)$,
have a much larger expressive power than the ones constructed from single-particle
orbitals.
In fact, such a \emph{generalized Slater determinant} (GSD) is in principle
sufficient to represent \emph{any} $N$-electron fermionic
wavefunction \cite{Pfau2020,Hutter2020}.
Nevertheless, the accuracy depends ultimately on the choice of the many-particle
functions $\phi_i(\br_j;\br_{/j})$, such that in practice using a small linear
combination of GSDs is advantageous.
The main innovation of the FermiNet ansatz is to express the wavefunction by
a linear combination of these generalized Slater determinants
\begin{equation}
  \label{eq:FN_Psi}
  \begin{aligned}
  \Psi_{FermiNet}(\br) &= \sum_{I=1}^{M} C_I \tilde{\Phi}_I(\br) \\
  &= \sum_{I=1}^{M} C_I \left(
  \mathcal{\hat{A}}\prod_{i=1}^N \phi_i^I(\br_i;/\br_i) \right)
  \end{aligned}
\end{equation}
using many-particle ``orbitals'' $\{\phi_i^I(\br_j;\br_{/j})\}$ represented by a
deep neural network.
Note that a different set of equivariant functions $\{\phi_i^I(\br_j;\br_{/j})\}$
is used for each GSD $\tilde{\Phi}_I$ in the superposition, which are indexed by
the superscript $I$.
Importantly, although \Cref{eq:FN_Psi} and \Cref{eq:FCI} look the same at first
sight, the former contains determinants which are able to describe complicated
many-body correlation effects through non-linear interactions between \emph{all}
electrons in each of the many-particle orbitals $\phi_i^I(\br_j;\br_{/j})$.
Hence, only a few determinants are sufficient to recover almost completely
the electron correlation.
The resulting architecture encoding the wavefunction is shown in \Cref{fig:FermiNet}.
To preserve the equivariance of the functions, FermiNet takes as input
electron-nuclear and electron-electron relative coordinates and
distances, and propagates through the network parallel streams of one-electron
and two-electron feature vectors $\mathbf{h}_i^l$ and $\mathbf{h}_{ij}^l$,
respectively.
Similar equivariant architectures have been adopted in other machine learning
approaches for computational chemistry, such as the SchNet neural
network \cite{Schutt2017,Schutt2018} discussed in chapter 12.
At each intermediate layer, these vectors are constructed by taking averages over
same-spin features, $\mathbf{g}_i^{l\sigma}$ with $\sigma \in \{\uparrow,\downarrow\}$,
which are concatenated to the distances and the output of the previous layer before
passing through a non-linear activation function. It is this repeated process through
the network that captures the correlation between the particles.
In the last layer, the determinants $\tilde{\Phi}_I(\br)$ are constructed, and their
linear combination produces the final wavefunction value for a given set of input
electronic and nuclear coordinates.
Owing to the non-linearity and complexity of the ansatz, the optimization
of the parameters relies on variational Monte Carlo, essentially following the
strategy outlined at the beginning of the Methods section.
However, in contrast to typical VMC wavefunction ansätze, FermiNet does not embed
any prior physical information, \textit{e.g.} the shape of the electron-electron
cusp usually modeled by the Jastrow factor. This fact leads to a more difficult
optimization procedure, such that pre-training is necessary to ensure convergence.
For instance, the many-electron orbitals can be pre-trained by minimizing
the least-square error to reference Hartree-Fock orbitals obtained in a
finite basis set calculation.

\textit{Quantum chemical applications.}\newline
FermiNet was tested on a variety of atomic and molecular systems \cite{Pfau2020}.
Ground state energies within chemical accuracy from the exact result
\cite{Chakravorty1993} and at most a few m$E_\text{h}$ from CCSD(T) at the
complete basis set (CBS) limit were obtained for first-row atoms, lithium through
neon.
Because FermiNet expresses the wavefunction directly in continuous space,
there is no one-particle basis set and thus the concept of basis set limit ceases
to exist in this framework. This is one of the major advantages with respect to
a formalism based on second quantization.
In comparison to a full CI wavefunction, the counterpart of the CBS extrapolation
would be the limit of the network to an infinite number of layers.
In small molecules containing two non-hydrogen atoms, FermiNet consistently
recovers more than $99\%$ of the correlation energy, while in larger molecular
systems --- methylamine, ozone, ethanol and bicyclobutane --- between $97\%$ and
$98\%$. For all these systems, it outperforms CCSD(T) in finite basis sets of
augmented quadruple and quintuple zeta quality, highlighting the superb accuracy
that can be reached with this ansatz.
The decline of the correlation energy recovered for larger systems could be
hinting at a problem with the size-extensive property of FermiNet. However, all
calculations were performed at a fixed network architecture, such that the
variational degrees of freedom were the same regardless of the size of the
systems considered. In contrast, traditional methods such as CCSD(T),
implicitly increase their flexibility through the basis sets; for a fixed
basis set quality, the number of basis functions depends on the size of
the molecule, such that for a size-extensive approach, the fraction of
correlation energy recovered should always be approximately the same.
The true power of FermiNet unfolds when considering systems with significant
strong correlation components. Here, the flexibility of the neural network
architecture allows for an accurate description of prototypical systems such
as the \ce{H_4} rectangle, the dissociation of \ce{N_2} and the \ce{H_10}
hydrogen chain. In all these cases, single-reference techniques such as coupled
cluster fail, and theoretical approaches considered state-of-the-art are
instead auxiliary-field quantum Monte Carlo \cite{Zhang2003} and multireference
configuration interaction.
Once again, FermiNet delivers results which are on-par with the best
methodologies available \cite{Pfau2020}.
For instance, in the dissociation of \ce{N_2}, the average error was about
5 milliHartree with respect to the experimental curve, comparable to the
highly accurate $r_{12}$-MR-ACPF results.
In the case of the hydrogen chain, stretching it from an internuclear
distance of 1 atomic unit up to 3.5, resulted in errors with respect to
MRCI+Q+F12 of less than 5 milliHartree, comparable to auxiliary-field
quantum Monte Carlo at the complete basis set limit.
The hydrogen chain was also used to investigate the effects of the network
architecture on the accuracy of FermiNet.
Here, it was found that the overall accuracy increases as more layers are
added to the network, suggesting that a deep architecture is advantageous
over a shallow, single-layer one. On the other hand, increasing the
the width of the layers also provided generally improved results.

\textit{Computational complexity.}\newline
Compared to conventional quantum chemical approaches, the evaluation of the
computational complexity of a neural-network-based ansatz is less
straightforward.
The total number of parameters in FermiNet is approximately given by
$\mathcal{O}(N_sn_h + Ln_h^2 + N_s^2M + N_sn_hM + M)$, where $N_s$ represents
the size of the system (number of electrons or atoms), $n_h$ the number of
hidden units, $L$ the number of layers and $M$ the number of determinants.
There are three terms that scale with the size of the system, the largest one
does so quadratically. The remaining terms are fixed for a given choice of
network architecture.
The evaluation of a single forward pass requires at most
$\mathcal{O}(N_s^3(M + n_h) + LN_s^2n_h^2)$ operations, where the first term
will dominate for larger system sizes at a fixed network architecture.
Evaluation of the local energy needed in the VMC framework requires the calculation
of the Laplacian, which scales with an extra factor of the system size.
The optimization procedure requires a matrix inversion, with the largest
component given by a term that scales as $\mathcal{O}(M^3N_s^6)$\footnote{
In the original article presenting FermiNet \cite{Pfau2020}, this
term is actually $\mathcal{O}((MN_eN_a)^3)$, where the number of atoms $N_a$
is not considered as "system size", which instead it is here (with $N_e$ being
the number of electrons). Hence the final 6th power scaling reported in this
chapter is in contrast to their asserted 4th power scaling due to the local
energy evaluation.}.
Overall, for a single optimization step, the scaling besides the matrix
inversion is a quartic power with respect to the system size (local energy
evaluation), which is to be compared to, for instance, the seventh power in
CCSD(T), and exponential for full CI.
While this analysis provides some theoretical ground on the computational
complexity behind FermiNet, in practice it is easier to provide actual
numbers for the performed calculations \cite{Pfau2020}.
In particular, for all the results discussed above, the architecture
included approximately $700000$ parameters, with resulting
training times (\textit{i.e.} wavefunction optimizations) between a few
hours for the smaller systems, up to a month for bicyclobutane using
8 to 16 GPUs.

\subsubsection{PauliNet}

Another successful example of deep neural network ansatz defined in
continuous space is PauliNet \cite{Hermann2020}.
While the functional form underlying FermiNet is completely general,
in the sense that it does not include any known physical feature of
the wavefunction besides the antisymmetry, PauliNet follows a more
traditional VMC approach, where deep neural networks are used to model
certain components of the ansatz.

\begin{figure}
  \centering
  \includegraphics[width=0.8\textwidth]{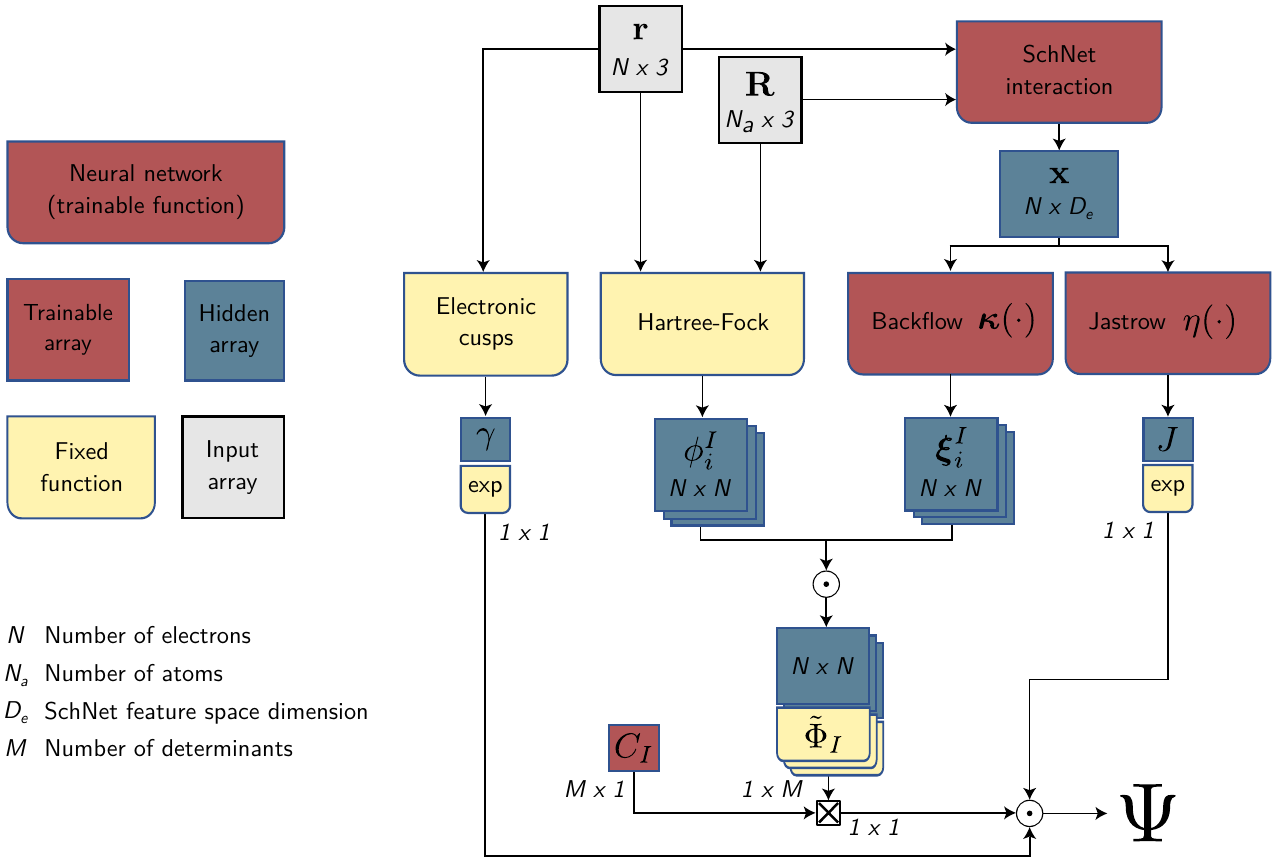}
  \caption{Simplified scheme depicting the structure of the PauliNet ansatz.}
  \label{fig:PauliNet}
\end{figure}

\textit{The ansatz.}\newline
PauliNet is a wavefunction ansatz of the Slater-Jastrow-backflow type,
and is given by
\begin{equation}
  \label{eq:PN_Psi}
  \begin{aligned}
  \Psi_{PauliNet}(\br)
  &= e^{\gamma(\br) + J(\br)} \sum_{I=1}^{M} C_I \tilde{\Phi}_I(\br) \\
  &= e^{\gamma(\br) + J(\br)} \sum_{I=1}^{M} C_I \left(
  \mathcal{\hat{A}}\prod_{i=1}^N \phi^I_i(\br_i) \xi^I_i(\br) \right)
  \end{aligned}
\end{equation}
There are essentially four components that make up $\Psi_{PauliNet}(\br)$.
The first two appear in the exponential factor in front of the sum:
$\gamma(\br)$ is a function that directly models the electronic cusp
of the wavefunction, while $J(\br)$ is a Jastrow factor, which captures
the short-range electron correlation effects.
The third component is the fixed linear superposition of (a few) Slater
determinants, which enforces the antisymmetry requirement of fermions.
The one-particle orbitals $\phi_i(\br_j)$ are standard Gaussian orbitals,
that are obtained by a Hartree-Fock or small complete active space SCF
calculation. The orbitals are multiplied by the fourth component, which
are the backflow functions $\xi_i^I(\br)$ (note the dependence on all
electronic coordinates).
As explained in the introduction to the VMC approach in the beginning of
the Methods section, such a transformation is crucial to improve the
nodal surface fixed by the Slater determinants.
However, contrary to \Cref{eq:backflow}, the backflow transformation in
PauliNet does not substitute the individual electronic coordinates in
$\phi_i(\br_j)$ by the corresponding backflow-transformed ones, but rather,
it directly multiplies the orbitals by many-electrons equivariant
functions $\xi_i^I(\br)$.
This choice leads to a simpler and more efficient optimization of the backflow
parameters.
The main innovation in the ansatz of \Cref{eq:PN_Psi} is the fact that the
Jastrow factor $J(\br)$ and the backflow transformation functions
$\xi_i^I(\br)$ are represented by deep neural networks, providing a very
flexible functional form for modeling both the electron correlation and
the nodal surface of the wavefunction.
For \Cref{eq:PN_Psi} to remain a valid wavefunction, the various components
need to satisfy different constraints.
To maintain the cusp conditions enforced by $e^{\gamma(\br)}$, the neural
networks parametrizing the Jastrow factor and the backflow transformation
are constructed cusp-less, that is, satisfying
\begin{align}
  \label{eq:PN_J_cusp}
  \nabla_{\mu_i} J(\br) \Big|_{\mu_i=\{\br_k,\bR_K\}} &= 0 \\
  \label{eq:PN_xi_cusp}
  \nabla_{\mu_i} \xi_i^I(\br) \Big|_{\mu_i=\{\br_k,\bR_K\}} &= 0
\end{align}
Furthermore, to preserve the antisymmetric nature of $\Psi_{PauliNet}(\br)$
imposed by the Slater determinants, $\gamma(\br)$ and $J(\br)$ are invariant
with respect to the exchange of pair of particles, that is
\begin{align}
  \label{eq:PN_gamma_inv}
  \gamma(\mathcal{P}_{ij}\br) &= \gamma(\br) \\
  \label{eq:PN_J_inv}
  J(\mathcal{P}_{ij}\br) &= J(\br)
\end{align}
where $\mathcal{P}_{ij}$ is the operator exchanging particles $i$ and $j$.
On the other hand, the backflow transformation functions $\xi_i^I(\br)$ are
equivariant, \textit{i.e.}
\begin{equation}
  \label{eq:PN_xi_equi}
  \mathcal{P}_{ij}\xi_i^I(\br) = \xi_j^I(\mathcal{P}_{ij}\br)
\end{equation}
The invariance and equivariance properties encoded in \Cref{eq:PN_J_inv}
and \Cref{eq:PN_xi_equi}, respectively, and the the many-particle dependence
of the electronic interactions are analogous to the requirements for learning
potential energy surfaces with neural networks.
Taking advantage of this similarity, another level of complexity is introduced
in the PauliNet architecture by transforming the electronic coordinates through
a modified version of SchNet \cite{Schutt2018}, before
they are fed to the Jastrow factor and backflow transformation.
In practice, SchNet projects each electronic coordinate onto a features
space of dimension $D_e \gg 3$, that encodes many-body correlations between
the particles. These high-dimensional representations, $\bx_i(\br)$, are then
used as input to the Jastrow and backflow functions
\begin{align}
  \label{eq:PN_Jastrow}
  J(\br) &= \eta \bigg( \sum_{i=1}^{N} \bx_i(\br) \bigg) \\
  \label{eq:PN_backflow}
  \bm\xi_i^I(\br) &= \bm\kappa \bigg( \sum_{i=1}^{N} \bx_i(\br) \bigg)
\end{align}
The functions $\eta(\cdot)$ and $\bm\kappa(\cdot)$ are modeled by deep neural
networks, with trainable parameters.
The electronic cusps, the Slater determinants with backflow transformed
coordinates and the Jastrow factor are then combined together to yield the
value of the wavefunction amplitude.
The overall PauliNet architecture is summarized in \Cref{fig:PauliNet},
which shows a simplified version of the ansatz (see \textcite{Hermann2020}
for a more comprehensive figure).
The sophisticated architecture of PauliNet reflects the many physical
components directly included in the ansatz. Importantly, while these
constrain its functional form, they do so without restricting its
representational power for modeling the wavefunction associated to
a quantum state. The various (fixed) components actually provide a blueprint,
upon which the deep neural networks describing the electronic embeddings,
the Jastrow factor and the backflow transformation provide sufficient
flexibility.
The optimization of the variational parameters underpinning PauliNet is
carried out in the framework of VMC. While the method to sample the
configurations and the technique to minimize the energy slightly differ
from the ones introduced above, the overall strategy is the same.

\textit{Quantum chemical applications.}\newline
PauliNet is able to recover between $97\%$ and $99.9\%$ of the correlation
energy of several atomic and diatomic systems, such as \ce{H_2}, \ce{LiH},
\ce{Be} and \ce{B} \cite{Hermann2020}.
This is achieved by training the neural networks for at most a few hours
on a single GPU.
This fact highlights an important difference with respect to the more
flexible FermiNet architecture. That is, the incorporation of the physical
components of a wavefunction in PauliNet (mean-field orbitals, electronic
cusp, and so forth) allows for a much faster optimization of the parameters.
The dependence of the energy with respect to the number of determinants
in \Cref{eq:PN_Psi} was assessed on four diatomic
molecules, \ce{Li_2}, \ce{Be_2}, \ce{B_2} and \ce{C_2} \cite{Hermann2020}.
It was found that PauliNet systematically recovers a larger fraction of the
correlation energy by increasing number of SDs, reaching the accuracy of
state-of-the-art diffusion Monte Carlo calculations with a much shorter
linear superposition compared to other VMC ansätze.
The use of a finite basis set for the mean-field orbitals appears to
introduce a basis set dependence in the calculation, which is a typical source
of error in traditional quantum chemical methods. Whereas CBS extrapolations
are required for the latter approaches, PauliNet is not sensitive to the
choice of basis set for the orbitals $\phi_i^I(\br_j)$ thanks to the deep
backflow transformation, which is able to compensate for the missing
flexibility in the basis \cite{Schatzle2021}.
Convergence to the fixed-node limit, \textit{i.e.} capturing all electron
correlation without modifying the position of the wavefunction nodes, can
also be reached by PauliNet. In particular, it was shown that for both
\ce{LiH} and \ce{H_4}, chemical accuracy can be quantitatively reached by
increasing the number of layers and their width in the deep Jastrow
factor \cite{Schatzle2021}.
Similarly to FermiNet, PauliNet is able to capture strong electron
correlation as well. For the challenging \ce{H_10} hydrogen chain,
$98.4\%$ of the correlation energy was recovered for both the equilibrium
and the stretched geometries, using a total of 16 determinants. The ansatz
performed extremely well also in combination with a single SD, with a
slightly less amount of $97.5\%$ of correlation energy captured.
While the calculations on small systems have shown that PauliNet can reach
chemical accuracy and surpass traditional quantum chemical methods, it
also scales to larger molecules. For instance, an investigation on
the automerization of cyclobutadiene, a molecule with 28 electrons,
provided results on-par with the best theoretical estimates for the
barrier, albeit with a smaller uncertainty interval.
Formally, the most expensive step in PauliNet is the evaluation of the
kinetic energy at the sample points, which scales as $\mathcal{O}(N^4)$,
with $N$ being a measure of the system size. This is the standard
computational complexity in variational Monte Carlo, and the presence
of the deep neural networks does not change it.

\subsection{Supervised machine learning of the wavefunction}

In the last two subsections we have seen examples of how machine learning
models can be used to represent the wavefunction ansatz directly, either
in Fock space or real space, and how these can be efficiently optimized
through variational Monte Carlo.
In this subsection we shall instead discuss a very different approach to
the quantum many-body problem, whereby the wavefunction is learned in
a supervised fashion.

\subsubsection{SchNOrb}

In electronic structure theory, the simplest description of a molecular
system is provided by a single SD wavefunction. This is constructed from
an antisymmetrized product of single-particle functions as discussed in
the previous sections (cfr. \Cref{eq:FN_SD}), which are typically
expanded in a linear combination of $N_{AO}$ local atomic orbitals (AOs) as
\begin{equation}
  \label{eq:schnorb_LCAO}
  \phi_i(\br) = \sum_{\mu=1}^{N_{AO}} c_{\mu i} \chi_\mu(\br)
\end{equation}
where $\br$ now denotes the Cartesian coordinates of a single electron.
Knowledge of the coefficients vectors $\{\bc_i\}$ and associated orbital
energies $\{\epsilon_i\}$ for a given basis set
$\{\chi_{\mu}(\br)\}_{\mu=1}^{N_{AO}}$ is sufficient to represent the
wavefunction and the total energy, and therefore to give access to molecular
properties of the modeled system as well.
Hence, it might be tempting to train a machine learning model to reproduce
these coefficients and energies for a prescribed atomic orbital basis set.
However, in practice, the training process is particularly difficult because
these are not smooth and well-behaved function of the nuclear coordinates,
displaying degenerate energies, changes of the orbital ordering and an
arbitrary dependence on the phase factor.
Instead, learning directly the representation of the Hamiltonian operator
in the atomic orbital basis, and the associated overlap matrix, leads
to a much better behaved problem that essentially contains the same
information.
In fact, given the Hamiltonian matrix $\bH$ and the overlap matrix $\bS$
between the basis functions $\chi_{\mu}(\br)$, the molecular orbital (MO)
coefficients $\bc_i$ and associated energies $\epsilon_i$ can be obtained
by a simple diagonalization of the following generalized eigenvalue problem
\begin{equation}
  \label{eq:schnorb_GEV}
  \bH \bc_i = \epsilon_i \bS \bc_i
\end{equation}
with the elements of $\bH$ and $\bS$ reading
\begin{align}
  \label{eq:schnorb_H}
  \bH_{\mu\nu} &= \braket{\chi_\mu|\hat{H}|\chi_\nu} \\
  \label{eq:schnorb_S}
  \bS_{\mu\nu} &= \braket{\chi_\mu|\chi_\nu}
\end{align}
In \Cref{eq:schnorb_GEV} and \Cref{eq:schnorb_H}, $\bH$ is either the
Fock matrix from Hartree-Fock theory or the Kohn-Sham matrix from density
functional theory.
Note that the idea to learn the Hamiltonian and overlap matrices in a given
basis is essentially the same as the approach taken in semi-empirical methods,
whereby \Cref{eq:schnorb_H,eq:schnorb_S} are parametrized against experimental
data or higher-level \textit{ab initio} calculations. The difference with
SchNOrb lies in how this parametrization is done. As we will see, using a
deep neural network leads to very accurate results.

\begin{figure}
  \centering
  \includegraphics[width=0.6\textwidth]{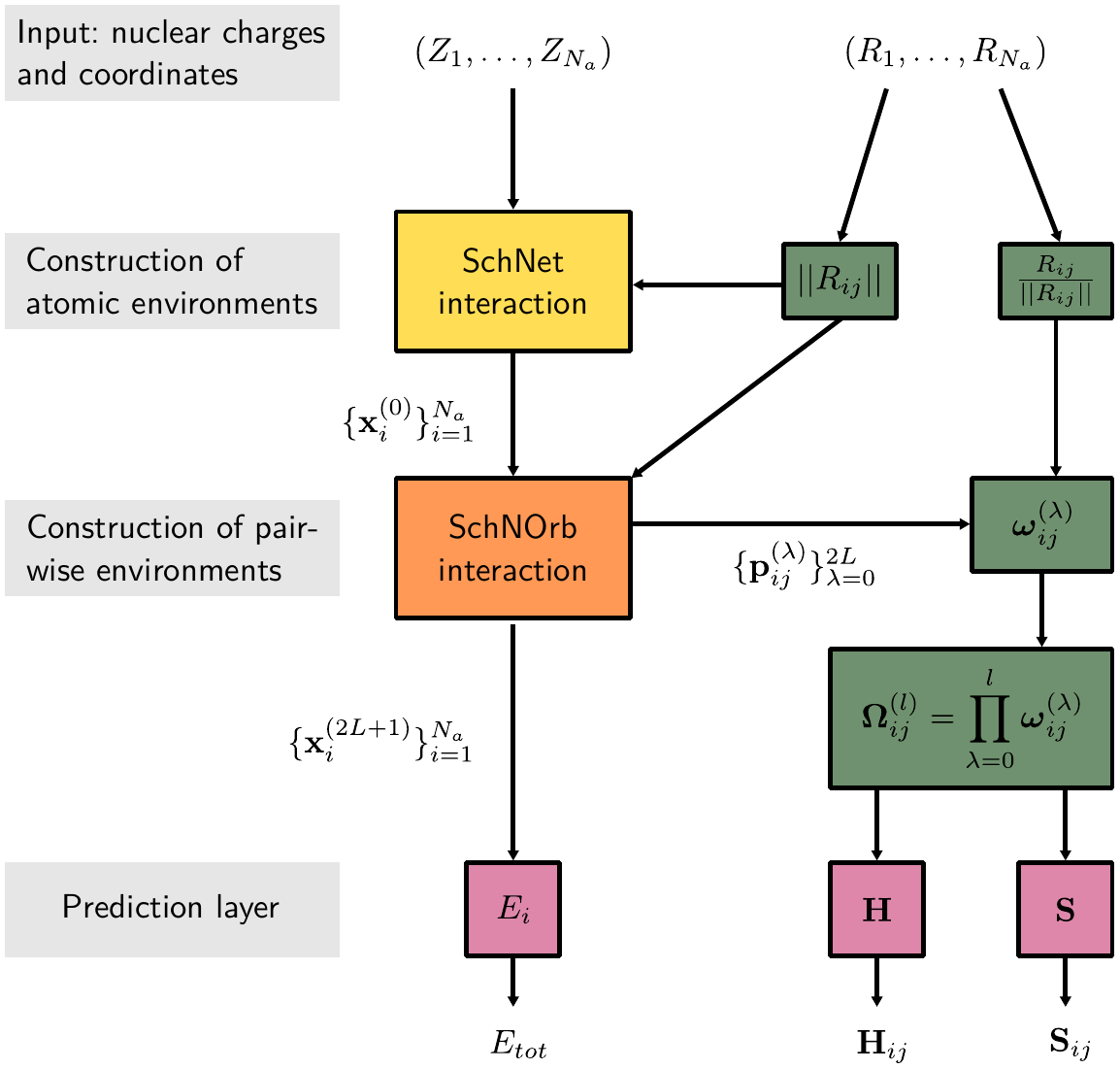}
  \caption{Simplified scheme depicting the SchNOrb deep neural network architecture.
  Green blocks are fixed functions, yellow and orange blocks are deep convolutional
  neural networks, and plum blocks are neural networks producing the final output.}
  \label{fig:schnorb}
\end{figure}

\textit{The SchNOrb neural network.}\newline
\emph{SchNOrb} is a deep learning framework in which the main idea is to
learn the representation of the Hamiltonian and overlap matrices from a large
set of reference calculations \cite{Schutt2019,Gastegger2020}.
This is achieved by training a deep convolutional neural network based on the
SchNet architecture \cite{Schutt2017,Schutt2018}.
The input consists of the atomic coordinates $(\bR_1,\ldots,\bR_{N_a})$ and
corresponding nuclear charges $(Z_1,\ldots,Z_{N_a})$.
Atomic representations are then constructed in a first stage following the
classical SchNet architecture, yielding
a set of high-dimensional feature vectors $\bx_i^{(0)}$.
The latter are passed on to the next stage, where a second deep
convolutional neural network, SchNOrb, generates pair-wise atomic features
\begin{equation}
  \label{eq:schnorb_Omega}
  \bm\Omega_{ij}^{(l)} = \prod_{\lambda=0}^{l} \bm\omega_{ij}^{(\lambda)}
\end{equation}
as products of $l$ symmetry-adapted polynomials of increasing order. This ensures
that the rotational symmetry of the local orbitals up to angular momentum $l$ can
be properly accounted for.
In each layer $\lambda$ of SchNOrb, the polynomials $\bm\omega_{ij}^{(\lambda)}$
depend on the pairwise interaction between the atomic environments $\bx_i^{(\lambda)}$
and $\bx_j^{(\lambda)}$, and the interatomic distance between atoms $i$ and $j$.
In this context, the polynomial coefficients $\bp_{ij}^{(\lambda)}$ making up
the $\bm\omega_{ij}^{(\lambda)}$'s can be thought of as the SchNOrb counterparts of
the linear coefficients of the orbital expansions in traditional quantum chemical
methods.
The pairwise features $\bm\Omega_{ij}^{(l)}$ constructed in this way are
then used to build the Hamiltonian $\mathbf{H}$ and overlap $\mathbf{S}$ matrix
representations.
Through the sequential passes in the layers of SchNOrb, the atomic environments
$\bx_i^{(\lambda)}$ are further refined and the output features vectors
$\bx_i^{(2L+1)}$ are used for the prediction of the total energy of the system $E_{tot}$.
A schematic representation of the SchNOrb architecture is shown in \Cref{fig:schnorb}.

\textit{Training and prediction.}\newline
The SchNOrb approach is a supervised learning algorithm. Reference data, that is,
the Hamiltonian and overlap matrices computed at a given level of theory (basis
set, method, functional), are generated by sampling the conformational space of
the molecule and by performing actual quantum chemical calculations.
The neural network is then trained with a combined regression loss given by
\begin{equation}
  \label{eq:schnorb_loss}
  l(\tilde{\bH},\tilde{\bS},\tilde{E},\bH,\bS,E,\mathbf{F}) =
  ||\bH - \tilde{\bH}||_F^2 +
  ||\bS - \tilde{\bS}||_F^2 +
  \rho||E - \tilde{E}||^2 +
  \frac{1-\rho}{N_{a}} \sum_{i=1}^{N_{a}} \bigg|\bigg|
  \mathbf{F}_i - \bigg(-\frac{\partial\tilde{E}}{\partial\br_i}\bigg)
  \bigg|\bigg|^2
\end{equation}
where the quantities with a tilde are the predicted values and $\rho$ determines
the trade-off between energy and forces.
The optimization is performed with standard procedures for training deep neural
networks, such as stochastic gradient descent.
For each molecule, several thousands geometries are necessary to properly sample
the conformational space and learn the correct rotational symmetries. For instance,
in the original work presenting SchNOrb \cite{Schutt2019}, 25000 conformations
were used for ethanol, malondialdehyde, and uracil.
Once trained, the performance of the neural network can be evaluated by computing
the mean absolute error between the prediction and reference calculations on a test
set containing new conformations.
An example of the accuracy that can be reached with SchNOrb is summarized in
\Cref{tab:schnorb_results}, for Hamiltonian and overlap matrices generated with
either Hartree-Fock or DFT in combination with the PBE exchange and correlation
functional and an atomic basis set including functions up to $d$ angular momentum.
\begin{table}
  \centering
  \caption{Mean absolute error of the SchNOrb prediction with respect
  to reference calculations of the Hamiltonian and overlap matrices, MO energies,
  occupied MO coefficients and total energies. The test set contained between 1500
  and 4500 new conformations.
  Data taken from \textcite{Schutt2019}.}
  \begin{tabular}{llcccccc}
  \toprule
  Molecule & Method & $\bH$ [meV] & $\bS$ & $\epsilon$ [meV] & $\phi_i$ & $E$ [meV] \\
  \midrule
  Water           & PBE & 4.5 & 7.91e-05 &  7.6 & 1.00 & 1.435 \\
  Ethanol         & HF  & 7.9 & 7.50e-05 & 10.6 & 1.00 & 0.378 \\
  Ethanol         & PBE & 5.1 & 6.78e-05 &  9.1 & 1.00 & 0.361 \\
  Malondialdehyde & PBE & 5.2 & 6.73e-05 & 10.9 & 0.99 & 0.353 \\
  Uracil          & PBE & 6.2 & 8.24e-05 & 47.9 & 0.90 & 0.848 \\
  \bottomrule
  \end{tabular}
  \label{tab:schnorb_results}
\end{table}
In all cases the accuracy is very good, with errors in the Hamiltonian matrix below
10 meV and in most cases below 1 meV for the total energy. Interestingly, the error
observed for the orbital energies (a derived property that is not directly learned)
is clearly distinct for occupied ($<20$ meV) and virtual orbitals ($\approx 100$ meV),
probably due to the fact that the latter are not strictly defined in the HF or
Kohn-Sham scheme.
Molecular properties are also accurate, with both dipole and quadrupole moments
reproduced with errors in the order of $0.055$ D and B, respectively.
Note that the accuracy of SchNOrb is bounded by the level of theory with which the
training data was generated. It would be desirable to have these in combination
with an accurate quantum chemical methods and, in particular, large basis sets.
However, larger bases imply an increased complexity and dimension of the
Hamiltonian and overlap matrices.
It was observed in this case, that while the prediction of the latter remained
accurate, the derived properties suffered from an increased error.
For instance, when the network was trained for ethanol in combination with a triple
zeta basis, a mean absolute error in the MO energies of 0.4775 eV was found,
highlighting the difficulty to learn the more complex representation \cite{Schutt2019}.
This error can be traced back to the diagonalization of the Hamiltonian matrix,
which in the larger basis accumulates the prediction error.
A solution to this issue is the projection of the calculations onto a optimized
minimal basis \cite{Gastegger2020}, which also has the advantage of shorter training
times due to the reduced dimensionality of the data.

\textit{SchNOrb applications}\newline
All the ML wavefunction approaches seen in the previous subsections
constitute novel ways to represent the complex functional form underlying a
quantum state. While these are inspired by machine learning models, their
practical application remains within a more traditional setting such as that
of variational Monte Carlo.
On the other hand, SchNOrb is an approach that follows the typical ML
paradigm more closley, thereby learning the relation between molecular
geometries and their quantum mechanical wavefunction representation from
large amount of data.
Because SchNOrb is bound to predict the wavefunction at an accuracy at most
comparable to the method used to generate the training data, the type of
applications targeted by this approach are different from the previous ones.
Even more so, considering that creation of the training set and the network
optimization requires a considerable amount of time.
For instance, for all molecules considered in the original work \cite{Schutt2019},
the training time was about 80 hours, and the creation of the training set took
from 65 hours for ethanol, up to 626 hours for uracil.
On the other hand, once the network is trained, the prediction is obtained in tens
of milliseconds, compared to seconds or minutes for the traditional quantum chemical
counterparts.
In this perspective, it is clear that SchNOrb is an ideal candidate to carry
out molecular dynamics simulations, at an accuracy well beyond that of classical
force fields, but at a comparable computational cost after training.
Other possible applications for SchNOrb could be to accelerate the convergence
of traditional SCF calculations, use of SchNOrb orbitals for post-HF methods,
or in inverse design, where a desired property could be optimized as a function
of the nuclear positions \cite{Schutt2019}.

\section{Case Studies}

In these cases studies we are going to use machine-learning-based methods to
solve the Schrödinger equation.
In the first case study, we will model the ground state wavefunction of a
quantum-mechanical particle in a one-dimensional box using Gaussian process
regression. This approach is similar in spirit to the Gaussian process state
ansatz presented in the Methods section, however, we will model the wavefunction
directly in real space rather than in Fock space.
In the second case study, we are going to compute the ground state energy of the
lithium hydride molecule using PauliNet, and analyze its dependence on the
basis set used to generate the orbitals.

\subsection{Particle in a Box}

Let us start with the theoretical background for this problem.
The Hamiltonian (in atomic units) to describe a quantum-mechanical particle
confined in a one-dimensional box of length $L$ is given by
\begin{equation}
  \label{eq:CS1_H}
  \hat{H} = -\frac{1}{2}\frac{\mathrm{d}^2}{\mathrm{d}x^2} + V(x)
\end{equation}
where the potential function $V(x)$ is
\begin{equation}
  \label{eq:CS1_V}
  V(x) = \begin{cases}
    0 , &0 < x < L \\
    \infty , &\textrm{otherwise}
  \end{cases}
\end{equation}
The exact ground state wavefunction for this Hamiltonian is known, and its
analytical form reads
\begin{equation}
  \label{eq:CS1_Psi}
    \Psi(x) = \begin{cases}
    \sqrt{\frac{2}{L}} \sin\left(\frac{\pi}{L}x\right) , &0 < x < L \\
    0 , &\textrm{otherwise}
  \end{cases}
\end{equation}
with the associated ground state energy
\begin{equation}
  \label{eq:CS1_E}
  E = \braket{\Psi|\hat{H}|\Psi} =
  \int \Psi^*(x)\hat{H}\Psi(x) \mathrm{d}x = \frac{\pi^2}{2L^2}
\end{equation}
The knowledge of the exact wavefunction will let us draw $N_{tr}$ random samples
$\{x^{(i)},\Psi(x^{(i)})\}_{i=1}^{N_{tr}}$ and create a training set from which
the underlying functional form of $\Psi(x)$ can be inferred from the data.

\subsubsection{Bayesian learning of the wavefunction with a Gaussian process}

\begin{figure}
  \centering
  \includegraphics[width=12cm]{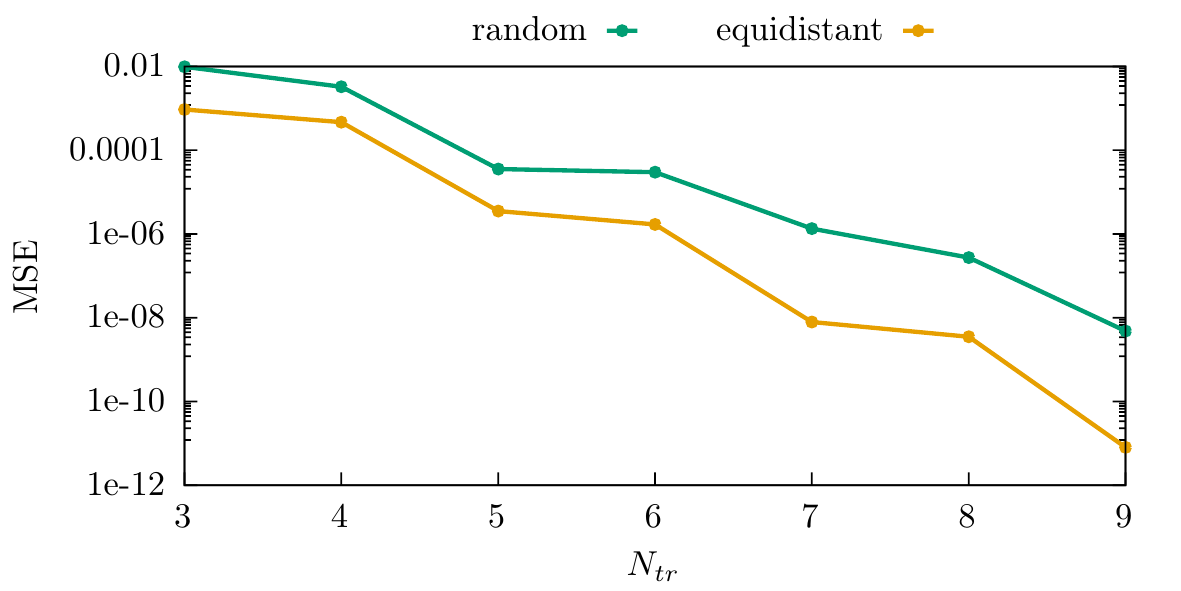}
  \caption{Convergence of the mean squared error in the wavefunction as a function
  of the training set size. The curve labeled ``random'' stands for the randomly
  generated training set, while the curve labeled ``equidistant'' stands for the
  training set generated on an equidistant grid of points.}
  \label{fig:CS1_error_conv}
\end{figure}

\begin{figure}
  \centering
  \includegraphics[width=12cm]{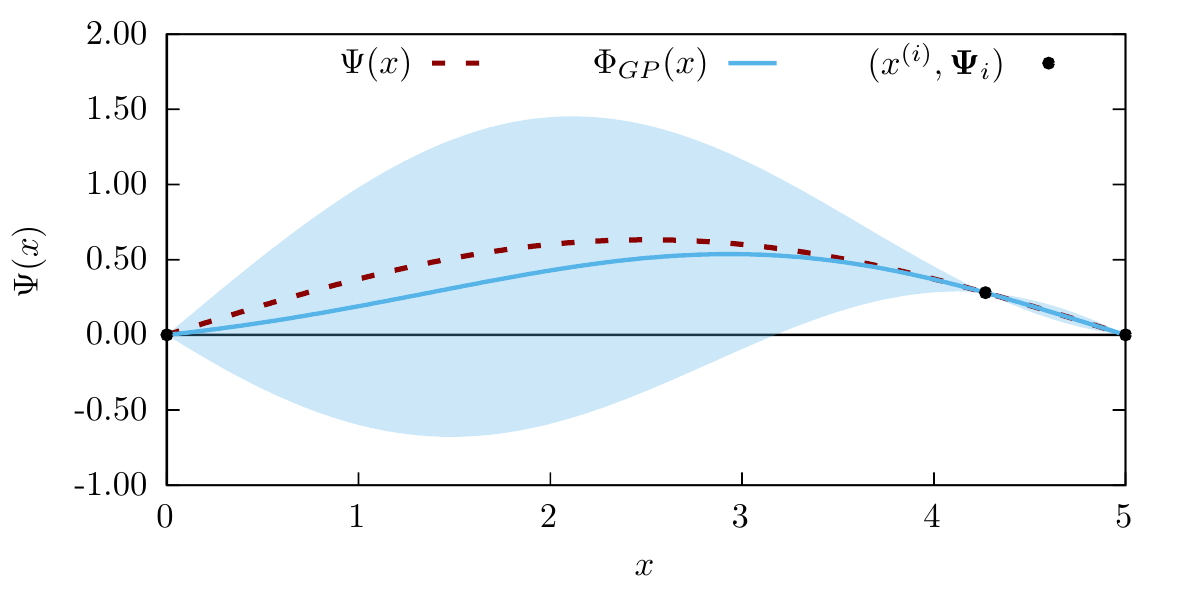}
  \caption{Gaussian process wavefunction trained on three points (black circles):
  a single randomly generated one within the box and two at the edges to ensure
  the proper boundary conditions. The shaded blue area is the 95\% confidence
  interval of the Gaussian process. The length of the box is $L=5$, the GP
  hyperparameters are $l=2$ and $\lambda = 0$.}
  \label{fig:CS1_GP_var}
\end{figure}

In this case study, we define a wavefunction ansatz as the mean of a
Gaussian process (GP)
\begin{equation}
  \label{eq:CS1_GPWF}
  \tilde{\Psi}(x) = \Phi_{GP}(x) = \begin{cases}
    \sum_{i=1}^{N_s} w_i k(x^{(i)},x) , &0 < x < L \\
    0 , &\textrm{otherwise}
  \end{cases}
\end{equation}
where we explicitly set to zero the wavefunction where the potential is infinite.
We will use the squared exponential kernel
\begin{equation}
  \label{eq:CS1_Gaussian_kernel}
  k(x,x') = \exp \left( -\frac{|x-x'|^2}{2l^2} \right)
\end{equation}
as this provides simple analytical forms for the derivatives that we need for
solving the electronic structure problem.
The hyperparameter $l$ is called the length-scale, and controls the locality
of the correlation between two points $x$ and $x'$. For a large value of $l$,
the kernel only correlates points very close to each other, and the opposite
is true for a small value of $l$. While the value of this hyperparameter can
in principle be optimized, we will not be concerned with that in this case
study, instead we will use a fixed value.
In the framework of Gaussian process regression, the vector of optimal weights
can be obtained in closed form with the following expression
\begin{equation}
  \bw = (\bK + \lambda \bm{1})^{-1} \bm\Psi
\end{equation}
where $\bK$ is the covariance matrix obtained by evaluating the kernel function
between all training points, $\bK_{ij} = k(x^{(i)},x^{(j)})$, $\bm\Psi$ is a
vector with the amplitudes of the exact wavefunction evaluated at all training
points, $\bm{\Psi}_i = \Psi(x^{(i)})$, and $\lambda$ is a small regularization
hyperparameter that controls the variance of the model at the training points.
For a value of $\lambda=0$, the Gaussian process will pass exactly through the
data used to train it, while for larger values it does not necessarily has to.
The wavefunction has to vanish at the edges of the box, that is
$\tilde{\Psi}(0) = \tilde{\Psi}(L) = 0$. Because the GP is an exact interpolator
for $\lambda = 0$, one way to enforce these boundary conditions
(bc) is to (always) include in the training set these two points at the
edges, $(x^{(1)}=0,\Psi(0)=0)$ and $(x^{(2)}=L,\Psi(L)=0)$.
This will ensure that $\Phi_{GP}(x)$ will take on those values at the boundaries.
We can now generate a few random training samples within the box and use these
together with the bc points to fit the ground state wavefunction.
To evaluate the accuracy of our model, we can compute the mean squared error
between the exact wavefunction and the ansatz at the test points
$\{\tilde{x}^{(i)}\}_{i=1}^{N_{pr}}$ (regularly spaced coordinates within the box)
\begin{equation}
  \label{eq:CS1_MSE}
  \text{MSE} = \frac{1}{N_{pr}}\sum_{y=1}^{N_{pr}}
  \big| \tilde{\Psi}(\tilde{x}^{(i)}) - \Psi(\tilde{x}^{(i)}) \big|^2
\end{equation}
As an exercise, plot the convergence of the error as a function of the
number of training points $N_{tr}$, using different strategies to generate them:
randomly draw them from a uniform distribution or place them equidistantly
within the box.
For this case study, you can use a box of length $L=5$, set the hyperparameter
of the Gaussian kernel to $l=2$ and the regularization parameter $\lambda = 0$.
To carry out these exercises, we provide a self-contained implementation in Python
based on the NumPy, SciPy and matplotlib libraries, presented as a jupyter notebook
that can be downloaded from
\href{https://github.com/stefabat/MLWavefunction}{github.com/stefabat/MLWavefunction}.
However, as an advanced exercise, you can try to write the program from scratch
yourself using your favorite programming language (hint: for Python use the NumPy
function \texttt{linalg.solve} to invert the covariance matrix).
You should obtain results similar to those shown in \Cref{fig:CS1_error_conv}.
One of the advantages of Gaussian processes is that they are defined with a
Bayesian inference framework. This allows to construct confidence intervals
about the model, that can tell us the uncertainty of the fitted function.
In particular, we can calculate the confidence for a prediction of the
wavefunction $\tilde{\Psi}(x)$ at a point $x$, based on the variance of
the GP model using the following formula
\begin{equation}
  \label{eq:CS1_conf_int}
  \sigma^2(x) = k(x,x) - \bk(x)^T (\bK + \lambda\bm{1})^{-1} \bk(x)
\end{equation}
where $\bk(x)$ is the vector function with elements $\bk_i(x) = k(x^{(i)},x)$,
with $x^{(i)}$ being the points of the training set.
We can now use \Cref{eq:CS1_conf_int} to plot the 95\% confidence interval for
each test point $\tilde{x}^{(i)}$, which approximately corresponds to twice the
variance, that is, $\tilde\Psi(\tilde{x}^{(i)}) \pm 2\sigma$.
This is shown in \Cref{fig:CS1_GP_var} for a randomly generated training set.
As we can see from the plot, the uncertainty is zero at the training
points because the GP passes right through them. On the other hand, maximum
variance is attained at regions that are farthest away from the underlying data
parametrizing the GP wavefunction.

\subsubsection{Variational optimization of the Gaussian process wavefunction}

\begin{figure}
  \centering
  \includegraphics[width=12cm]{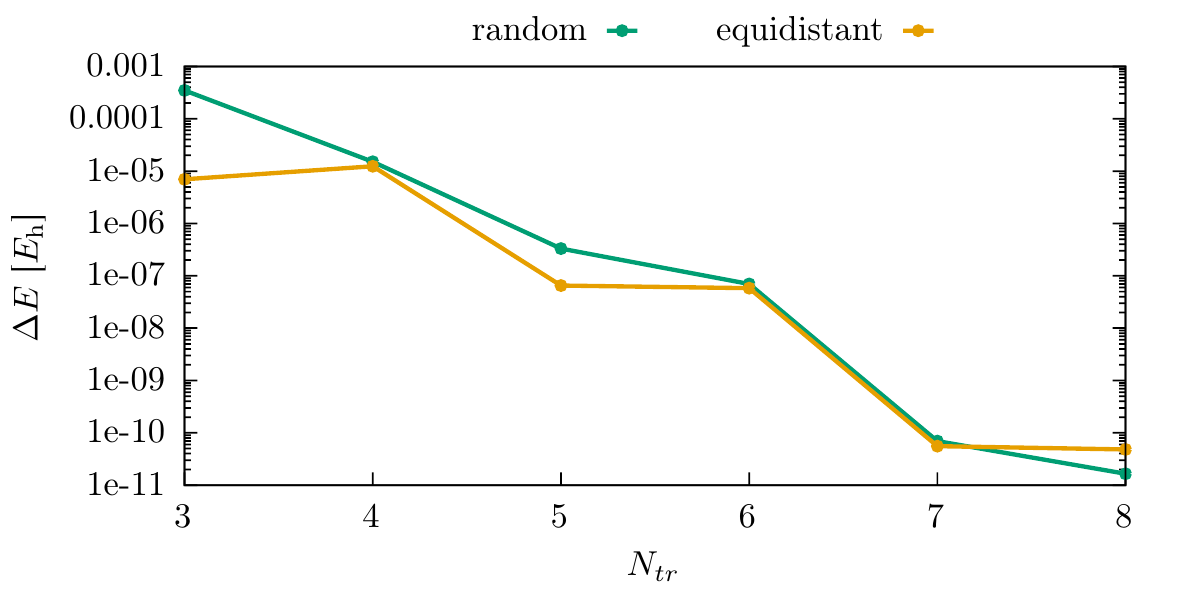}
  \caption{Ground state energy difference between the variationally optimized
  GP wavefunction and the exact solution for different training set sizes.
  The curve labeled ``random'' stands for the randomly generated training set,
  while the curve labeled ``equidistant'' stands for the training set generated
  on an equidistant grid of points.}
  \label{fig:CS1_GP_var_opt}
\end{figure}

In the second part of the first case study, we will optimize the Gaussian
process wavefunction variationally, obtaining the optimal weights without
the need to know the wavefunction \textit{a priori}.
To do so, it is first convenient to slightly change our ansatz of
\Cref{eq:CS1_GPWF}, by explicitly incorporating the boundary conditions.
This can be done by multiplying the GP with a function that satisfies the
constraints, that is
\begin{equation}
  \label{eq:CS1_bc}
  f(x) = x(L-x)
\end{equation}
yielding a new ansatz
\begin{equation}
  \label{eq:CS1_GPWF_f}
  \tilde{\Psi}(x) = \begin{cases}
    \Phi_{GP}(x)f(x) , &0 \le x \le L \\
    0 , &\textrm{otherwise}
  \end{cases}
\end{equation}
Note the explicit definition of $\Phi_{GP}(x)f(x)$ at the boundaries
$x = 0$ and $x = L$ included (in contrast to \Cref{eq:CS1_GPWF}).
It is easy to test that $\tilde\Psi(x) = \Phi_{GP} x(L-x)$ at these
points is exactly zero.
The variational optimization of \Cref{eq:CS1_GPWF_f} requires first an
expression for the energy expectation value, \textit{i.e.}
\begin{equation}
  \label{eq:CS1_H_expval}
  \tilde{E} = \frac{\Braket{\tilde{\Psi}|\hat{H}|\tilde{\Psi}}}
  {\Braket{\tilde{\Psi}|\tilde{\Psi}}} =
  \frac{\Braket{\tilde\Psi|
  -\frac{1}{2}\frac{\mathrm{d^2}}{\mathrm{d}x^2} + V(x)
  |\tilde\Psi}}{\Braket{\tilde{\Psi}|\tilde{\Psi}}}
\end{equation}
Substitution of $\tilde{\Psi}(x)$ with \Cref{eq:CS1_GPWF_f} results in
the following double sum for the Hamiltonian expectation value in the
numerator
\begin{equation}
  \label{eq:CS1_H_matrix}
  \sum_{i=1}^{N_s} \sum_{j=1}^{N_s} w_i^* w_j
  \Braket{k(x^{(i)},x)f(x)|\hat{H}|k(x^{(j)},x)f(x)} =
  \sum_{i=1}^{N_s} \sum_{j=1}^{N_s} w_i^* w_j H_{ij} =
  \bw^T \bH \bw
\end{equation}
and similarly for the overlap in the denominator
\begin{equation}
  \label{eq:CS1_S_matrix}
  \sum_{i=1}^{N_s} \sum_{j=1}^{N_s} w_i^* w_j
  \Braket{k(x^{(i)},x)f(x)|k(x^{(j)},x)f(x)} =
  \sum_{i=1}^{N_s} \sum_{j=1}^{N_s} w_i^* w_j S_{ij} =
  \bw^T \bS \bw
\end{equation}
The evaluation of \Cref{eq:CS1_H_matrix} requires the
second-order derivative of the $k(x^{(i)},x)f(x)$ with respect to $x$.
Fortunately, the simple form of the squared exponential kernel we have selected
admits an analytical expression, which is left as an exercise to derive.
Furthermore, because the wavefunction ansatz is zero everywhere but
within the box, we can restrict the domain of integration for the matrix
elements to go from $0$ to $L$, where $V(x) = 0$.
In principle, the matrix elements can be obtained in closed form, however,
for simplicity we will use numerical quadrature to perform the integration
(hint: if you are writing the program from scratch, you can use the SciPy
function \texttt{integrate.quad}).
The last step required for the variational optimization of $\tilde\Psi(x)$
is to minimize the energy with respect to the weights $w_i$ of the GP.
These act as the variational degrees of freedom of our ansatz.
While the minimization of $\tilde{E}$ can be performed with common methods
used in machine learning, such as (stochastic) gradient descent, for our
simple example, it is more convenient to equate the derivative of the energy
with respect to the weights, and explicitly solve the resulting generalized
eigenvalue problem
\begin{equation}
  \frac{\mathrm{d}\tilde{E}}{\mathrm{d}w_i} = 0
  \; \text{for} \; i=1,\ldots,N_{tr}
  \iff \bH\bw = \tilde{E}\bS\bw
\end{equation}
This is completely analogous to the way in which configuration interaction
wavefunctions are optimized in quantum chemistry.
Note that, as a byproduct, the diagonalization of $\bH$ will also return
the weights associated to the excited states of the particle.
As an exercise, study the energy convergence of the GP wavefunction for
increasing number of training points $N_{tr}$. As before, you can generate
the points randomly or equidistantly within the box.
Note that a training point in this context can be understood as the center
of a new basis function used to represent $\tilde\Psi(x)$.
You should get results similar to the ones shown in \Cref{fig:CS1_GP_var_opt}.
As we can see, for such a simple system, there is not much difference in the
energy convergence between the randomly generated training set and the
equidistant one. In line with the GPS ansatz discussed in the Methods section,
for a small training set, \textit{e.g.} with $N_{tr} = 3$, the training point
within the box is very important for the accuracy. The randomly generated one
yields a GP wavefunction with a significantly worse energy than with the
training point placed at the center of the box.

\subsection{Ground state energies from PauliNet}

In this second case study we will compute the ground state energy of lithium
hydride using PauliNet \cite{Hermann2020}, investigate its dependence on the
basis set used to generate the initial orbitals and compare it to highly-correlated
CCSD(T) calculations.
First, install the DeepQMC package following the installation guide at
\href{https://deepqmc.github.io/installation.html}{deepqmc.github.io}.
Note that the full package with all dependencies is more than 1GB of data.
The installation of DeepQMC will also pull the quantum chemical library PySCF,
which we will use to perform the coupled cluster calculations.

We provide a jupyter notebook written in python that walks you through this
second case study, that can be downloaded from
\href{https://github.com/stefabat/MLWavefunction}{github.com/stefabat/MLWavefunction}.
However, you can also try to carry out the steps described
below by yourself as an exercise.

\subsubsection{PauliNet training}

We are going to use the default PauliNet architecture and change the basis set
to the correlation-consistent family \cite{DunningJr1989}, starting from cc-pvdz
to cc-pv5z.
The first step is to define a \texttt{Molecule} object (for example named
\texttt{LiH}) with the geometry of LiH at an internuclear distance of
$3.0141132$ Bohr, and a total zero charge and spin.
Then, the PauliNet neural network can be initialized for a given basis set,
say cc-pvdz, from a Hartree-Fock calculation using the function
\texttt{from\_hf(LiH, basis='cc-pvdz')}, meaning that the ansatz contains
a single Slater determinant (cfr. \Cref{eq:PN_Psi}) and uses the HF orbitals.
After initialization, the network can be trained using the function
\texttt{train(net)}, where \texttt{net} is the neural network object returned
from the initialization procedure.
This is the most expensive part of the calculation, however, the training
parameters can be slightly modified to decrease the computation time, and still
obtain reasonably optimized network weights (with an uncertainty of around
1 m$E_\text{h}$).
We suggest you to use \texttt{n\_steps = 500}, \texttt{batch\_size = 500}
and \texttt{epoch\_size = 20}.
The last step to obtain the ground state energy of LiH is to call the function
\texttt{evaluate(net)}. Similarly to \texttt{train(net)}, \texttt{evaluate(net)}
accepts optional parameters defining the number of Monte Carlo sweeps and
the length of the Markov chain sampled at each step. For these two, we suggest
the following values, \texttt{n\_steps = 400} and \texttt{sample\_size = 800}.
Save the obtained energy and the estimated error, and repeat the same process
for the other basis sets.
The choice of training parameters was such that the optimization does not take
too long time.
If you want to investigate the effects of the training procedure on the final
energy and uncertainty, as an advanced exercise you can modify the values of
\texttt{n\_steps} and \texttt{batch\_size} in the \texttt{train(net)} function
(hint: modify one value at a time to understand their role).

\subsubsection{CCSD(T) calculations}

In this part we are going to perform CCSD(T) calculations in combination with
the four basis sets used above. You can use PySCF for performing these
calculations, and they should take significantly less time than the PauliNet
ones. PySCF has a pretty extensive user guide available online at
\href{https://pyscf.org/}{https://pyscf.org/}, with many examples.
We suggest you to have a look at it if you get stuck at any point.
The first step is again to define the lithium hydride molecule, this time in
a data-structure that PySCF understands.
This can be done with the function \texttt{gto.M(\ldots)}, which accepts a number
of arguments similar to the construction of the \texttt{Molecule} object
for the DeepQMC package.
Then, you can create a restricted HF instance with \texttt{rhf = scf.RHF(mol)},
where \texttt{mol} is the molecule you just created (hint: make sure to set the
correct basis set with \texttt{mol.basis = 'cc-pvdz'} and build it with
\texttt{mol.build()}).
The \texttt{rhf} object can be used to create the coupled cluster instance
with \texttt{cc = cc.CCSD(rhf)}.
The energy for each method can then be obtained by calling \texttt{.kernel()}
on the HF and coupled cluster objects, and \texttt{.ccsd\_t()} for the
perturbative triples correction.
Save all these energies and repeat the process for all basis sets.
If you have a powerful computer, as an additional exercise you can try to
compute the full CI energy, however, you will probably be able to get it at
most with the triple zeta basis set (hint: for such a small molecule, the
CCSD(T) and FCI energies are very similar).

\subsubsection{Comparison of PauliNet and CCSD(T)}

\begin{figure}
  \centering
  \includegraphics[width=9cm]{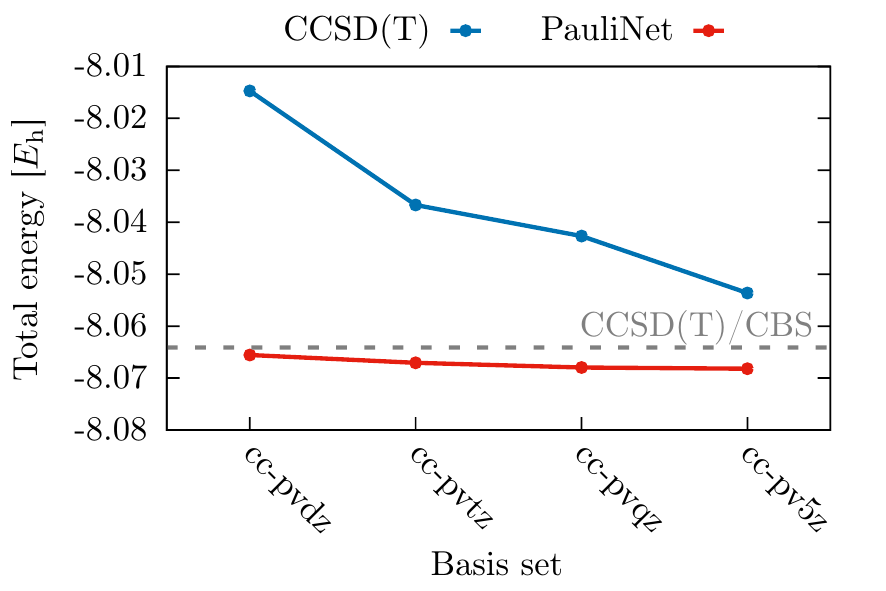}
  \caption{Ground state energy convergence of LiH as a function of
  the basis set for CCSD(T) and PauliNet. The dashed gray line
  corresponds to the CCSD(T) energy at the complete basis set limit.}
  \label{fig:CS2_energies}
\end{figure}

If you correctly performed all calculations you should obtain a plot similar
to that of \Cref{fig:CS2_energies}.
As we can see, CCSD(T) has a very marked dependence on the basis set size.
This is a well known shortcoming for all methods defined in Fock space, which
suffer from the so-called basis set truncation error.
We can estimate what is the CCSD(T) energy at the complete basis set limit
by using an extrapolation scheme.
Because the Hartree-Fock energy and the correlation energy behave differently
as a function of the basis set, we need to extrapolate them separately, using
two different extrapolation schemes.
For the Hartree-Fock energy, we can use the following formula
\begin{equation}
  E_{RHF}^{(X)} = E_{RHF}^{CBS} + A \exp\big( -\alpha \sqrt{X} \big)
\end{equation}
where $A$, $\alpha$ and $E_{RHF}^{CBS}$ are fitting parameters which we
can find with the calculations with triple, quadruple and quintuple zeta
basis sets, and $X$ is the cardinal number associated to the basis set:
2 for double, 3 for triple, and so forth.
Similarly, the correlation energy (note, only the correlation energy without
including the Hartree-Fock part) can be extrapolated according to
\begin{equation}
  E_{CC}^{CBS} = \frac{X^B E_{CC}^{(X)} - Y^B E_{CC}^{(Y)}}
  {X^B - Y^B}
\end{equation}
where $Y = X + 1$ and $B$ can be either obtained again by fitting, or
set to the value of $B=3$.
The plot shown in \Cref{fig:CS2_energies} provides us important insight.
First, PauliNet is very insensitive to the choice of basis set. This
point was discussed in more detail in the subsection dedicated to ML methods
in real space, and is empirically demonstrated here.
Second, the correlation energy recovered by PauliNet is more than
that from CCSD(T). Considering that PauliNet provides results which are
intrinsically at the complete basis set limit regardless from the orbitals
used, it implies that the difference between PauliNet and CCSD(T)/CBS
is due to missing many-body correlation not captured by the latter.
However, as you may have experienced by following this case study, this
superb accuracy comes at much more expensive price.
A comprehensive investigation of PauliNet with respect to both the basis
set limit and the correlation energy limit is presented in \textcite{Schatzle2021}.

\section{Conclusions and Outlook}

In this chapter we have seen several unsupervised machine learning approaches
to solve the Schrödinger equation. The main advantage is clear, machine learning
offers very flexible functional forms which are able to model highly complex
many-body correlation effects governing the quantum world.
In contrast, traditional quantum chemical methods rely on wavefunction ansätze
expressed in a finite number of electronic configurations, which typically span
only a fixed and delimited region of an otherwise exponentially large space of
functions.
While this provides an intuitive framework to describe a quantum state and allows
for straightforward optimization schemes, a fixed parametrization significantly
constrains the variational freedom of the wavefunction.
The risk is to miss important correlation effects outside the range modeled
and include instead redundant information of unimportant ones.
Machine learning wavefunction ansätze such as Boltzmann machines, Gaussian
process states and deep neural networks, start with a different premise.
The modeled correlation features are not enforced \textit{a priori}, instead
the wavefunction has complete freedom to take on whatever functional form
best describes the quantum state.
This has several advantages, for instance there is no more distinction between
single-reference and multi-reference, or, in the case of real space approaches,
the results do not depend on a basis set anymore.
The hard task is then shifted to the optimization of the variational parameters.

For the most part, the ansätze discussed in this chapter are based on
highly non-linear maps between input (electronic coordinates or many-body
configurations) and output (the associated wavefunction amplitude).
This gives rise to two inter-related challenges. On the one hand,
the evaluation of expectation values, such as the energy, requires the
integration of a high-dimensional function that does not admit an analytical
expression.
On the other hand, the optimization of the parameters underlying these ansätze
is a complicated and computationally demanding minimization problem.
The general framework in which these two challenges have been addressed is
variational Monte Carlo.
This has worked very well to provide compelling evidence that machine learning
wavefunction methods are extremely powerful approaches to tackle the quantum
many-body problem, but certainly leaves room for improvement for the future.
For instance, due to the sharply peaked distribution of wavefunction amplitudes,
algorithms that better sample the configuration space (\textit{i.e} with a
higher acceptance rate) are needed; see for example \textcite{Wu2021}.
For methods that can admit analytically tractable expressions, the stochastic
framework could also be abandoned in favor of deterministic optimization
schemes. This is the case of Gaussian process states, and perhaps highlights
one of the advantages of kernel methods over neural networks.

Other interesting developments can be envisioned in the realm of real space
approaches. While Fock space methods have overwhelmingly dominated the first
wave of machine learning wavefunction, the superb accuracy reached by deep
neural networks and their intrinsic independence from a single-particle
basis are extremely appealing features. Here, one of the main issues remains
the implementation of the antisymmetry property into the wavefunction.
Generalized Slater determinants provide a neat solution to this problem,
but other routes are possible and being explored \cite{Inui2021}.
Furthermore, kernel-based methods to model wavefunctions in real space are
still missing. We have already seen a glimpse of their potential in the first
case study, but it is expected that more concrete examples will appear
soon.
Furthermore, it is worth exploring the connections between the different
representations provided by neural networks and non-parametric models
expressed in real space, to those defined in Fock space. As it is often
the case, different perspectives on the same method (or representation)
can provide additional insight in its mathematical and physical
interpretations.

Unsupervised machine learning algorithms are not the only ones that have
been developed to model the wavefunction. The SchNOrb neural network has
shown an alternative route to deal with the Schrödinger equation.
Directly learning the wavefunction is not strictly required, and it might
instead be convenient to train a neural network to predict the
Hamiltonian representation in a finite basis.
In this case, the extensive supervised training required is a computationally
demanding task: first, reference data needs to be generated and second,
the model needs to learn it comprehensively.
However, SchNOrb offers an appealing complementary approach to the
unsupervised wavefunction ansätze discussed in this chapter, that has very
interesting potential applications, \textit{e.g.} in molecular dynamics
and inverse chemical design.

To conclude, the combination of machine learning and wavefunction theory
holds great promise for the future, and constitutes a field that is still in
its infancy. Many important developments are thus expected in the future.

\section*{Acknowledgments}

The author is thankful to Roland Lindh for insightful discussions about Gaussian
process regression and valuable feedback on the chapter.
Furthermore, the author acknowledges the Swiss National Science Foundation
(SNSF) for the funding received through the Postdoc Mobility fellowship.

\printbibliography

\end{document}